\documentclass[journal,twocolumn,10pt]{IEEEtran}
\IEEEoverridecommandlockouts 
\usepackage{epsfig,latexsym}
\usepackage{float}
\usepackage{indentfirst}
\usepackage{amsmath}
\usepackage{amssymb}
\usepackage{times}
\usepackage{subfigure}
\usepackage{pifont}
\usepackage{psfrag}
\usepackage{cite}
\usepackage{url}
\usepackage{lastpage}
\usepackage{bm}
\usepackage{color}
\usepackage{fancyhdr}
\usepackage[center]{caption2}
\usepackage{balance}
\usepackage{fancyhdr,lastpage}
\usepackage{hyperref}
\usepackage{epstopdf}

\begin{document}
	\title{Rydberg Atomic Quantum Receiver Empowered Hybrid Non-orthogonal Multiple Access }
	\author{\IEEEauthorblockN{Xinya Shen,~Zheng Chu,~\emph{Member, IEEE}, 
			David Chieng,~\emph{Senior Member, IEEE} \\ Chiew Foong Kwong,~\emph{Senior Member, IEEE},~Qihao Peng,~\emph{Member, IEEE},~Pei Xiao,~\emph{Senior Member, IEEE}\\
			Theodoros Tsiftsis,~\emph{Senior Member, IEEE}, and Jianyue Zhu,~\emph{Member, IEEE}
		}
		\thanks{Xinya Shen, Zheng Chu, David Chieng, and Chiew Foong Kwong are with the Department of Electrical and Electronic Engineering, University of Nottingham Ningbo China, Ningbo 315100, China, and also with the Next Generation Internet of Everything Laboratory, University of Nottingham Ningbo China, Ningbo 315100, China (Email: xinya.shen@nottingham.edu.cn; andrew.chuzheng7@gmail.com; David.Chieng@nottingham.edu.cn; Chiew-Foong.Kwong@nottingham.edu.cn)}
		\thanks{Qihao Peng and Pei Xiao are with the 5GIC \& 6GIC, Institute for Communication Systems (ICS), the University of Surrey, Guildford, GU2 7XH, UK. (Email: q.peng@surrey.ac.uk; p.xiao@surrey.ac.uk)}
		\thanks{Theodoros Tsiftsis is with the Department of Electrical and Electronic Engineering, University of Nottingham Ningbo China, Ningbo 315100, China, is also with the Department of Informatics and Telecommunications, University of Thessaly, 35100 Lamia, Greece (Email: tsiftsis@uth.gr)}
        \thanks{Jianyue Zhu is with the College of Electronic and Information Engineering, Nanjing University of Information Science and Technology, Nanjing 210044, China (Email: zhujy@nuist.edu.cn)}
		}
	\vspace{-0.5in}
	\maketitle 
	\thispagestyle{empty}
	\begin{abstract}
		In this paper, we propose a novel uplink Hybrid Non-Orthogonal Multiple Access (NOMA) architecture assisted by Rydberg Atomic Quantum Receivers (RAQRs) to improve energy efficiency by addressing the physical constraints of atomic sensing. We formulate a joint optimization problem encompassing phase measurements and multi-user power allocation to minimize the total transmit power. This problem is particularly challenging due to the non-linear amplitude-only observations of atomic vapor cells. To decomposes the resulting non-convexity, we develop a hierarchical three-layer decomposition optimization framework that systematically decouples the global problem. The inner layer employs a customized Expectation Maximization-Gerchberg-Saxton (EM-GS) algorithm to resolve phase ambiguity and estimate equivalent channel gains. The middle layer derives closed-form optimal power allocation solutions via Karush-Kuhn-Tucker (KKT) conditions. The outer layer utilizes linear Minimum Mean Square Error (MMSE) detection for high-reliability signal recovery. Simulation results demonstrate the superiority of the proposed framework over conventional RF-based Hybrid NOMA. Notably, the atomic architecture achieves approximately a 93\% reduction in total power consumption compared to Orthogonal Multiple Access (OMA) baselines while maintaining strict backward compatibility, thereby laying a solid foundation for practical deployment in next-generation green communication networks.
	\end{abstract}
	\IEEEpeerreviewmaketitle
	\setlength{\baselineskip}{1\baselineskip}
	\newtheorem{definition}{Definition}
	\newtheorem{fact}{Fact}
	\newtheorem{assumption}{Assumption}
	\newtheorem{theorem}{Theorem}
	\newtheorem{lemma}{Lemma}
	\newtheorem{corollary}{Corollary}
	\newtheorem{proposition}{Proposition}
	\newtheorem{example}{Example}
	\newtheorem{remark}{Remark}
	\newtheorem{algorithm}{Algorithm}
	\section{Introduction}
	\label{sec:introduction}
	The main goals of sixth-generation (6G) wireless networks are to achieve: global coverage, full spectrum utilization, comprehensive application support, and robust security \cite{you2021towards, sharma2019toward}. Within this expanded scope, 6G must support the massive deployment of energy-constrained devices, ranging from low-power sensors to autonomous industrial terminals. However, accommodating such a sheer volume of connections poses a critical bottleneck due to the physical limitations of available bandwidth and the finite battery life of IoT devices. Therefore, a fundamental challenge in realizing this vision lies in designing highly efficient communication frameworks that can serve a massive number of users while rigorously optimizing scarce spectral and energy resources.
	
	To address these constraints, innovations are needed in both access protocols and physical layer hardware. On the protocol front, the rigidity of conventional orthogonal access is increasingly ill-suited for the sporadic, massive nature of IoT traffic \cite{sharma2019toward}. This has driven the shift toward Hybrid NOMA, which integrates non-orthogonal schemes with legacy systems to enhance spectral efficiency and flexibility \cite{liu2022evolution,ding2024next,sun2025energy}. In parallel, the uplink performance of energy-starved devices is often restricted by the thermal noise floor and sensitivity limits of classical radio-frequency (RF) receivers. To detect ultra-weak signals beyond these limits, Rydberg Atomic Quantum Receivers (RAQRs) have emerged as a revolutionary solution in quantum sensing \cite{gong2025rydberg,degen2017quantum,chen2025harnessing}. Recent research has coalesced around these two pillars: the algorithmic evolution of multiple access flexibility \cite{ding2024design,ding2025study,ding2024hybrid} and the hardware advancement of receiver sensitivity \cite{anderson2021atomic,artusio2022modern,cui2025rydberg}.
	
	\subsection{Literature Review}
    \subsubsection{Evolution Towards Flexible and Efficient Hybrid Multiple Access}
    To address the spectral efficiency limitations of OMA, power-domain NOMA was introduced. NOMA enables multiple users to share time-frequency resources via superposition coding and successive interference cancellation (SIC) \cite{vaezi2019multiple, liu2022evolution}. While identified as a key enabler for IMT-2030 \cite{3GPP_36_211_2019}, the rigid implementation of NOMA causes compatibility issues. Consequently, Hybrid NOMA has emerged as a critical evolution, designed to allow the dynamic coexistence of OMA and NOMA within a unified framework \cite{ding2024next}. This architecture facilitates opportunistic access, where a weak user transmits in a time slot allocated to a strong user, provided the interference remains manageable thereby enabling adaptive multi-timeslot resource allocation \cite{sun2025energy, ding2024design, ding2025study}.
    
    This flexibility renders Hybrid NOMA highly effective for IoT scenarios characterized by sporadic traffic. Recent studies have optimized protocols to accommodate bursty uplink transmissions \cite{ding2024hybrid}, focusing on joint resource allocation to maximize sum rates. Furthermore, the integration of Hybrid NOMA with simultaneous wireless information and power transfer (SWIPT) and backscatter communications is being actively explored. These approaches jointly optimize energy harvesting and data transmission, offering a pathway to support battery-free devices \cite{she2021tutorial, ding2025backcom, huang2025energy}. Collectively, these advancements position Hybrid NOMA as a leading candidate for Next-Generation Multiple Access (NGMA) in 6G, balancing high throughput with massive connectivity and energy sustainability.
    
    \subsubsection{RAQRs}
    While advanced multiple access schemes improve spectral utilization, uplink performance for energy-constrained devices remains bounded by receiver sensitivity. Conventional RF chains are inherently limited by thermal noise and finite linearity, creating a bottleneck for ultra-weak signal detection. Quantum sensing offers a paradigm shift to overcome these classical limitations \cite{gong2025rydberg, degen2017quantum}.
    
    Among quantum platforms, RAQRs have shown significant potential for wireless communications \cite{gyongyosi2019survey, wang2022quantum}. By leveraging the large electric dipole moments of Rydberg atoms and Electromagnetically Induced Transparency (EIT) \cite{saffman2010quantum}, RAQRs convert incident RF signals into optical readouts with quantum-limited noise performance \cite{anderson2021atomic}. This mechanism provides distinct advantages, including sensitivity approaching the standard quantum limit, immunity to thermal noise, wideband frequency tunability, and high linearity \cite{cui2025rydberg, liu2025sa}.
    
    The development of RAQRs has progressed rapidly from theoretical physics to practical communication systems. Early experiments successfully demonstrated the reception of amplitude- and frequency-modulated signals using atomic vapors \cite{fan2015atom, meyer2018digital, yuan2023rydberg}. A significant milestone was the realization of the atomic superheterodyne receiver, which drastically improved sensitivity and validated the technology for precision metrology \cite{jing2020atomic}. Recent comprehensive reviews provide detailed insights into the operational principles of these receivers and their expanding utility in electric field sensing \cite{simons2021rydberg, fancher2021rydberg, yuan2023quantum}.

    On the theoretical front, significant progress has been made in bridging quantum physics with communication theory. Equivalent baseband channel models have been established for both single-input single-output (SISO) and multiple-input multiple-output (MIMO) configurations of RAQRs, allowing for the quantification of performance gains over conventional receivers \cite{cui2025towards, gong2025equivalent}. Crucially, specific detection and estimation algorithms have been developed to address the unique “amplitude-only” observation characteristic of certain RAQR implementations, which inherently results in phase ambiguity \cite{cui2025towards, gong2025equivalent}. Research has further accelerated into complex communication scenarios, investigating RAQR-specific precoding for MIMO systems \cite{peng2025ris,gong2025rydbergMIMO, cui2024mimo}, hardware-constrained channel estimation \cite{xu2025channel}, and multi-band operation for spectrum aggregation \cite{cui2025rydberg2,wang2025multi}. Additionally, applications in spatial signal processing, such as direction-of-arrival estimation, have highlighted the transition of RAQRs toward practical, ultra-sensitive communication hardware \cite{gong2025rydberg2,chen2025polarization}.
    
	\subsection{Motivation and Contribution}
    Although significant research has advanced NOMA for next-generation networks and explored the high-sensitivity capabilities of RAQRs, their synergistic integration for communication systems remains a major gap. Specifically, leveraging RAQRs' unique wideband and ultra-high-sensitivity to enhance NOMA's flexibility and reliability in dynamic scenarios is unexplored \cite{meyer2020assessment}. Furthermore, their exceptional detection capability offers a novel physical-layer solution to address the persistent challenge of error propagation in SIC, promising reliable interference management at significantly lower power \cite{chen2025harnessing}. Moreover, while energy efficiency is a key NOMA objective, existing strategies focus on algorithmic power allocation. Phase-coherent RAQR arrays offer a novel method to significantly reduce transmit power, which is essential for energy-limited 6G IoT applications\cite{santamaria2022comparison}. Thus, it is imperative to exploit a NOMA framework harnessing RAQR to simultaneously address flexibility, reliability, and energy efficiency.
	
	To bridge this gap, this paper conceives a novel uplink Hybrid NOMA system assisted by an RAQR at the base station and develops a comprehensive optimization framework. To the best of the authors’ knowledge, the exploitation of RAQR-empowered Hybrid NOMA systems remains largely unexplored. 
    
\emph{The main contributions of this paper are summarized as follows:}
	
\begin{itemize}
    \item We first propose a RAQR-based Hybrid NOMA uplink network, motivated by the need to enhance system flexibility through dynamic access scheme adaptation and improve energy efficiency in power-constrained scenarios, explicitly incorporating the critical hardware constraint of amplitude-only observations. To evaluate the energy-efficient performance metric under this framework, we formulate the joint optimization problem that encompasses phase alignment, multi-user power allocation, and OMA compatibility requirements to minimize the total transmit power.
    
    \item Next, a hierarchical three-layer solution strategy is proposed to address the non-convexity of the optimization problem while achieving reliable interference cancellation. This method systematically decomposes the intractable global problem into solvable sub-problems: (i) an \textit{inner layer} for phase retrieval and channel estimation utilizing a customized EM-GS algorithm; (ii) a \textit{middle layer} for resource allocation that derives closed-form power solutions via Karush-Kuhn-Tucker (KKT) conditions; and (iii) an \textit{outer layer} for signal detection employing an online MMSE scheme.
    
    \item We validate the superiority of the proposed architecture over conventional RF-based benchmarks through extensive simulations. The results demonstrate significant gains in both sum rate and energy efficiency across diverse channel conditions. This confirms the system's potential in enabling sustainable, high-performance connectivity for next-generation energy-constrained IoT devices, while leveraging the high receiver sensitivity of RAQR arrays to achieve reliable weak signal detection at reduced transmit power.
\end{itemize}
	
	The remainder of this paper is summarized as follows. Section~\ref{section:system model} presents the system model. Section~\ref{section: framework} details the proposed three-layer optimization framework. Simulation results are presented in Section~\ref{sec:simulation}. Finally, the paper is concluded in Section~\ref{section:Conclusion}. 
	
	\subsection{Notations}
    For clarity, the following notation is adopted throughout this paper. Italic letters denote variables, bold letters denote vectors. $\mathbb{R}$ and $\mathbb{C}$ represent the real and complex number fields, respectively. The operators $|\cdot|$ and $\|\cdot\|$ denote the element-wise modulus and the Euclidean norm of a vector, respectively. $\mathcal{R}\{\cdot\}$, $\mathcal{J}\{\cdot\}$, and $\angle(\cdot)$ denote the real part, imaginary part, and phase of a complex variable, respectively. The superscripts $(\cdot)^T$ and $(\cdot)^*$ stand for the transpose and the optimal value under current conditions. Moreover, $\hat{x}$ and $\tilde{x}$ denote the estimated and transformed values of variable $x$, while the superscript $(\cdot)^\text{opt}$ refers to the optimally phase-aligned equivalent real-valued channel. $\mathcal{Q}(\cdot)$ represents the Pulse Amplitude Modulation (PAM) demapping operation. Finally, $\mathcal{CN}(\mu, \Sigma)$ is a circularly symmetric complex Gaussian distribution with mean $\mu$ and covariance $\Sigma$. $\circ$ represents the Hadamard (element-wise) product.
	
\section{System model and problem statement}\label{section:system model}
\subsection{System Model}
\begin{figure}[H]
	\centering
	\includegraphics[width=1\linewidth]{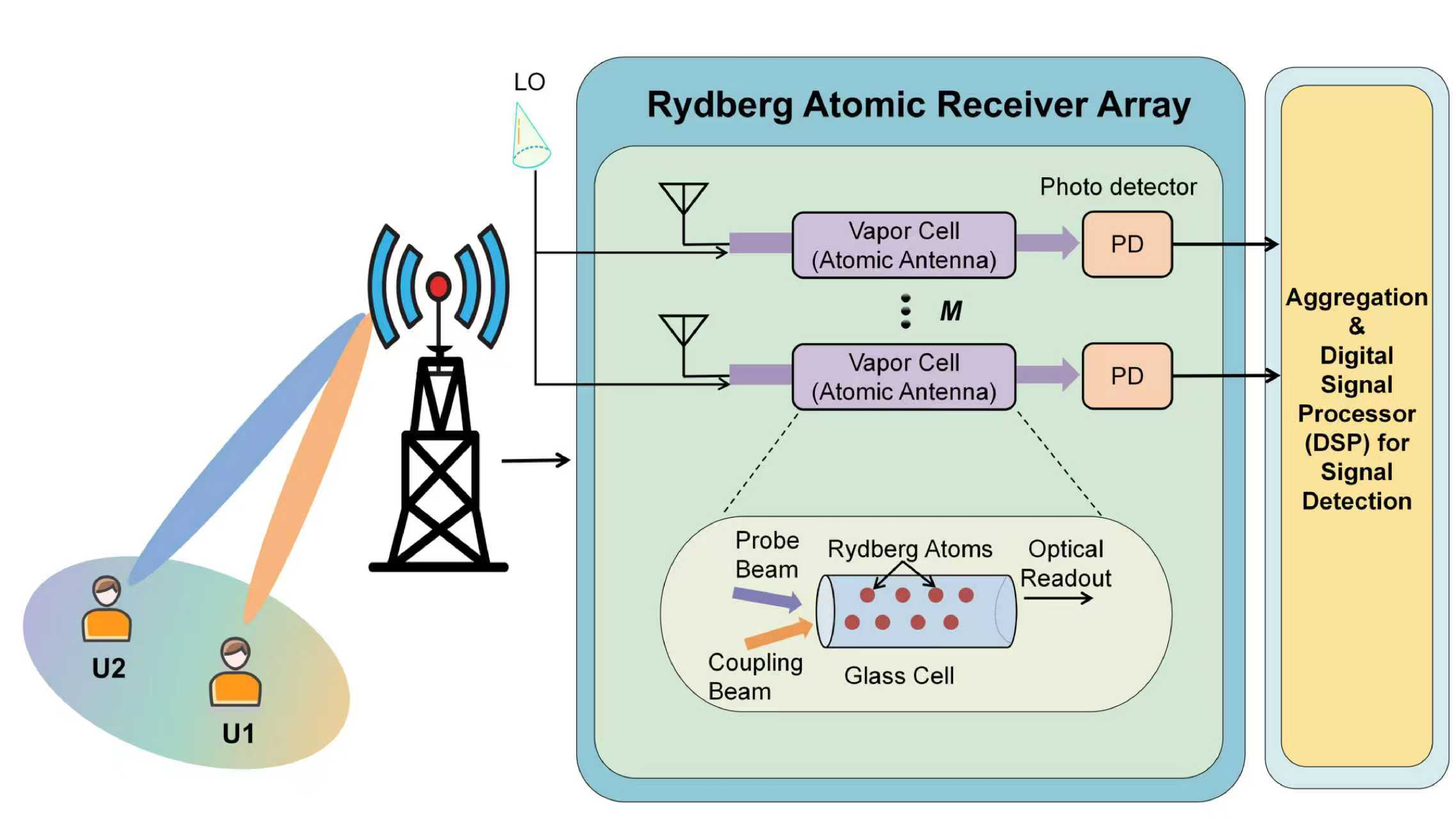}
	\caption{A two-user Hybrid NOMA with RAQR}
	\label{fig:system model}
\end{figure}
\begin{figure}[H]
	\centering
	\includegraphics[width=0.7\linewidth]{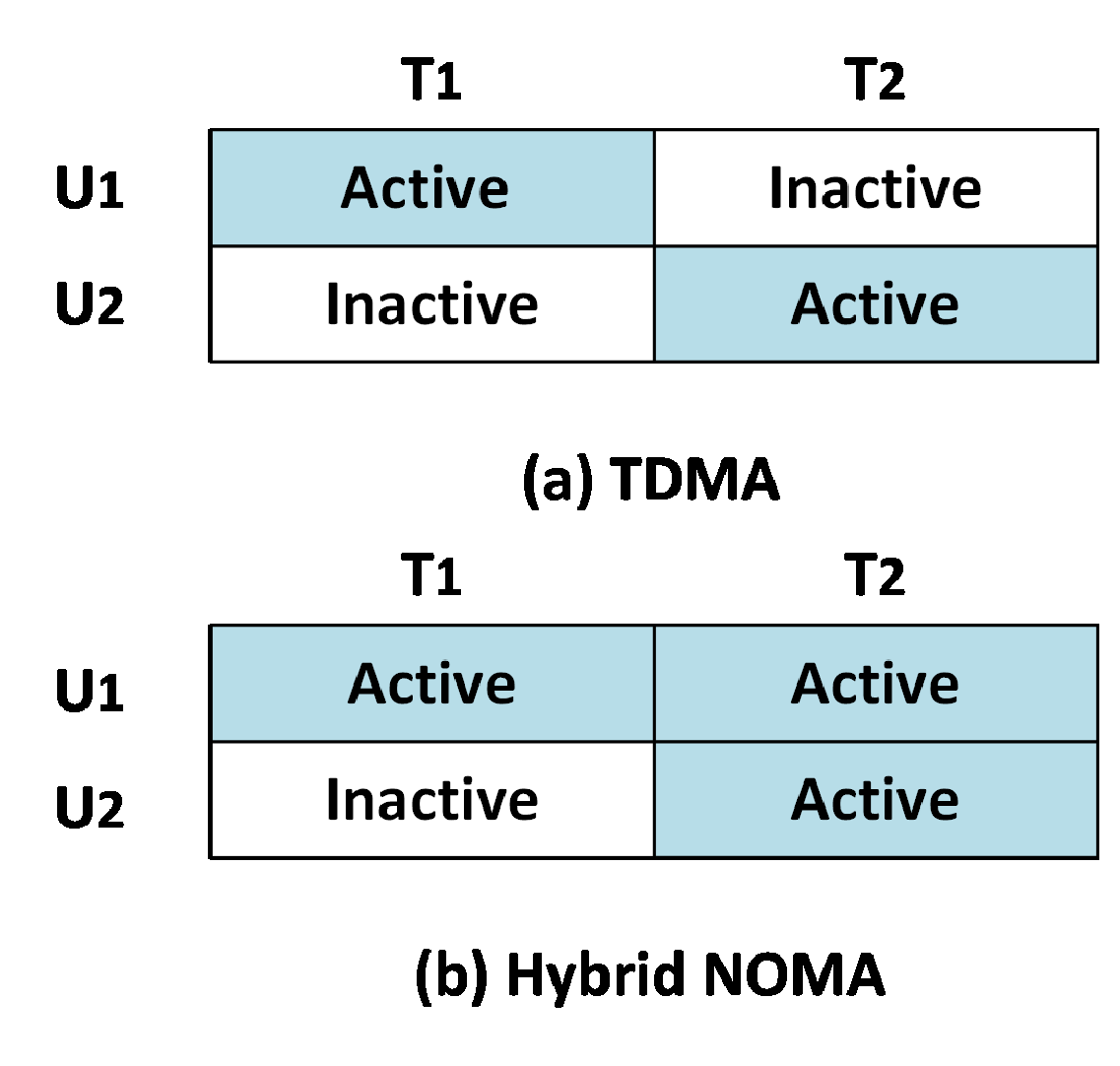}
	\caption{A two-user Hybrid NOMA transmission scheme}
	\label{transmission strategies}
\end{figure}
In this section, we consider an uplink network where two users, $\mathrm{U}_k$ ($k \in \{1,2\}$), are served by a single base station. Detailed physical principles of the RAQR can be found in \cite{cui2025towards}; here, we focus on the equivalent communication model. As illustrated in Fig. \ref{fig:system model}, the base station replaces the conventional RF front-end with an $M$-element RAQR. This array consists of $M$ vapor cells acting as atomic antennas, each filled with Rydberg atoms and coupled to a dedicated photodetector (PD). The optical readouts from these PDs are aggregated and processed digitally for signal detection.

Under this system architecture, the transmission protocol employed by the users is indicated in Fig. \ref{transmission strategies}. For comparison, Fig. \ref{transmission strategies} (a) presents the conventional OMA ( Time Division Multiple Access, TDMA) baseline, where resources are strictly orthogonal: time slot $\mathrm{T}_1$ is dedicated exclusively to $\mathrm{U}_1$, and $\mathrm{T}_2$ to $\mathrm{U}_2$, with each user transmitting at a baseline power $P_k^{\mathrm{OMA}}$. While, the Hybrid NOMA framework adopted in this study, based on \cite{ding2025study}, is illustrated in Fig. \ref{transmission strategies} (b). This scheme enables full-time resource sharing through dynamic role assignment. In time slot $\mathrm{T}_1$, $\mathrm{U}_1$ functions as the primary (strong) user, while $\mathrm{U}_2$ is superimposed as the secondary (weak) user. Conversely, in time slot $\mathrm{T}_2$, the roles are reversed, with $\mathrm{U}_2$ becoming the primary user and $\mathrm{U}_1$ the secondary user. Under this scheme, the transmit power for user $k$ in time slot $t$ is variable and denoted by $P_{k,t}$.

Building upon the aforementioned network architecture and transmission protocol, let us define the complex baseband signal transmitted by user $k$ in time slot $t$ as $s_{k,t} = s_k e^{j\psi_k}$, where $s_k$ is the real-valued PAM information symbol and $\psi_k$ denotes the initial phase. The complex signal vector $\mathbf{y}_t$ at the output of the Rydberg atomic quantum receiver (RAQR) is modeled as 
\begin{align}
	\mathbf{y}_t = \sum_{k=1}^2 \mathbf{h}_{k,t} s_{k,t} + \mathbf{b} + \mathbf{n}_t,
    \label{eq:rx-signal}
\end{align}
where $\mathbf{h}_{k,t} \in \mathbb{C}^{M \times 1}$ denotes the channel vector from user $U_k$ to the $M$-element atomic array, $\mathbf{b} \in \mathbb{C}^{M \times 1}$ is the local-oscillator reference signal, and $\mathbf{n}_t \sim \mathcal{CN}(0,\sigma^2\mathbf{I})$ is the additive noise. Equation~\eqref{eq:rx-signal} represents the equivalent complex baseband model derived from the Rydberg electromagnetically induced transparency (EIT) principle. Under small-signal conditions, the down-converted optical field is a linear superposition of the users' symbols $\{s_{k,t}\}$ weighted by the channel vectors $\{\mathbf{h}_{k,t}\}$, plus the deterministic LO term $\mathbf{b}$. The photodetector output thus follows the linear model in~\eqref{eq:rx-signal}, analogous to a conventional RF front-end. In practical Rydberg receivers, the dominant microscopic noise mechanism is signal-dependent shot noise rather than purely additive Gaussian noise \cite{wang2020tight}, but for analytical tractability we adopt the widely used circularly symmetric complex Gaussian approximation based on the central limit theorem, i.e., $\mathbf{n}_t \sim \mathcal{CN}(0,\sigma^2\mathbf{I})$, which enables the subsequent rate analysis and power optimization.

Unlike the traditional RF receivers that directly measure this complex signal $\mathbf{y}_t$, the RAQR in this work can only measure the amplitude of the received signal $\mathbf{y}_t$. Coherent detection of both amplitude and phase could alternatively be realized by employing a pair of lock-in amplifiers \cite{meyer2018digital,simons2019rydberg}; investigating such receiver architectures for Hybrid NOMA is an interesting direction for future work.
\begin{align}
\mathbf{z}_{t} = \lvert \mathbf{y}_{t} \rvert 
= \left\lvert \sum_{k=1}^{2}\mathbf{h}_{k,t}s_{k,t} + \mathbf{b} + \mathbf{n}_{t} \right\rvert 
\in \mathbb{R}^{M\times1},
\label{eq:Observation model}
\end{align}
where the amplitude-only observation $\mathbf{z}_{t}$ is directly linked to the Autler--Townes (AT) splitting interval of the underlying Rydberg EIT resonance, since the magnitude of the probe transmission at each atomic element is (after calibration) approximately proportional to the AT splitting induced by the total RF field $\sum_{k=1}^{2}\mathbf{h}_{k,t}s_{k,t} + \mathbf{b} + \mathbf{n}_{t}$. This non-linear amplitude relationship poses a significant challenge for signal detection. However, the signal vector $\mathbf{z}_{t}$ can be linearized by ensuring phase coherence between the received signal and the strong LO. Specifically, provided that the phase condition $\angle(\mathbf{h}_{1,t}s_{1,t} + \mathbf{h}_{2,t}s_{2,t}) \approx \angle \mathbf{b}$ holds, the signal vector $\mathbf{z}_{t}$ is reformulated by the small-signal approximation, which is given by
\begin{align}
\mathbf{z}_{t} \approx \lvert \mathbf{b} \rvert 
+ \mathbf{h}_{1,t}^\mathrm{opt}s_{1,t} 
+ \mathbf{h}_{2,t}^\mathrm{opt}s_{2,t} 
+ \mathbf{\hat{n}}_t,
\label{eq:simplified form of the signal}
\end{align}
where $\mathbf{h}_{k,t}^\mathrm{opt}$ denotes the equivalent real-valued optical channel for user $k$ in time slot $t$, and $\mathbf{\hat{n}}_t$ represents the effective noise. It is worthwhile noted that achieving precise phase alignment transforms the EIT/AT-based non-linear observation in~\eqref{eq:Observation model} into the linear model in~\eqref{eq:simplified form of the signal}, facilitating efficient user signal recovery and interference cancellation.

\subsection{Problem Statement}
Following the linearized observation model derived in \eqref{eq:simplified form of the signal}, we define the effective channel gain $G_{k,t}(\boldsymbol{\psi})$ as the squared Euclidean norm of the equivalent real-valued channel vector and express it as $G_{k,t}(\boldsymbol{\psi}) = \|\mathbf{h}_{k,t}^\mathrm{opt}\|^2 = \sum_{m=1}^{M}(h_{m,k,t}^\mathrm{opt})^{2}$, where it quantifies the aggregate signal combining gain provided by the $M$-element atomic MIMO array under a specific phase alignment $\boldsymbol{\psi}$. Since the base station prioritizes decoding the strong user $\mathrm{U}_{1}$ in the uplink, the presence of the weak user $\mathrm{U}_{2}$ would degrade the signal-to-interference-plus-noise ratio (SINR) of $\mathrm{U}_{1}$. To tackle this issue, we aim to prevent $\mathrm{U}_{2}$ from interfering with $\mathrm{U}_{1}$ in this time slot. To proceed, let us denote the achievable data rate of $\mathrm{U}_{k}$ at time slot $\mathrm{T}_{t}$ as $R_{k,t}$. Hence, at the time slot $\mathrm{T}_{1}$, the two users' data rates $ R_{1,1} $ and $ R_{2,1} $ are given, respectively, by 
    \begin{align}\label{eq:R in T1}
		R_{1,1} = \log_{2}\left(1+\frac{P_{1,1}\cdot G_{1,1}(\boldsymbol{\psi})}{\sigma^{2}}\right), ~
		R_{2,1} =  0.
\end{align}
In the time slot $T_{2}$, NOMA transmission is employed to enhance system performance. By implementing SIC at the base station, the signal of the strong user $U_{1}$ is decoded first. Thus, the achievable data rates $ R_{1,2} $ and $ R_{2,2} $ are expressed, respectively, as
\begin{subequations}\label{eq:R in T2}
    \begin{align}
		R_{1,2} &=\log_{2}\left(1+\frac{P_{1,2}\cdot G_{1,2}(\boldsymbol{\psi})}{P_{2,2}\cdot G_{2,2}(\boldsymbol{\psi})+\sigma^{2}}\right), \label{eq:R in T21} \\
		R_{2,2} &=\log_{2}\left(1+\frac{P_{2,2}\cdot G_{2,2}(\boldsymbol{\psi})}{\sigma^{2}}\right)
		\label{eq:R in T22}.
\end{align}
\end{subequations} 
As indicated by \eqref{eq:R in T1} and \eqref{eq:R in T2}, the legacy users achieve data rates identical to those of the OMA baseline within their respective allocated slots, thereby strictly guaranteeing backward compatibility. Consequently, the total achievable data rate for $U_1$ is given by $R_1 = R_{1,1} + R_{1,2}$, whereas $U_2$ transmits exclusively in the second time slot, yielding $R_2 = R_{2,2}$. Furthermore, $R$ denotes the system target rate, which is determined by the specific quality-of-service (QoS) requirements.

On the basis of the aforementioned discussion, we formulate the optimization problem by minimizing the total power consumption of both users, subject to the achievable data rate requirement of each user as well as the Frobenius norm constraint of the composite effective channel phase at each time slot. Mathematically, this optimization problem can be expressed as follows:
\begin{subequations}\label{eq:Oriproblem}
	\begin{align}
		\min_{P_{k,t}\geq0, \boldsymbol{\psi}}\quad&\sum_{k=1}^2\sum_{t=1}^2 P_{k,t} \label{eq:TotalPower}\\
		\text{s.t.} \quad & R_{1} \geq R, \label{eq:R1 constraint}\\
		& R_{2} \geq R, \label{eq:R2 constraint}\\
		&\left\|\mathfrak{I}\left\{\mathbf{H}_{t}^{\mathrm{eq}}(\boldsymbol{\psi})\right\}\right\|_{F}^{2} \leq \epsilon, \quad \forall t. \label{eq:phase constraint}
	\end{align}
\end{subequations}
where the effective channel gain, denoted as $G_{k,t}(\boldsymbol{\psi}) = \bigl\Vert \mathfrak{R}\{\mathbf{h}_{k,t} \circ e^{j\boldsymbol{\psi}}\} \bigr\Vert^{2}$, explicitly depends on the phase vector $\boldsymbol{\psi}$. Also, the composite effective channel matrix is defined as $\mathbf{H}_{t}^{\mathrm{eq}}(\boldsymbol{\psi}) = [\mathbf{h}_{1,t} e^{j\psi_1}, \mathbf{h}_{2,t} e^{j\psi_2}]$. In problem \eqref{eq:Oriproblem}, the objective function \eqref{eq:TotalPower} is the total transmit power consumption; the constraints \eqref{eq:R1 constraint} and \eqref{eq:R1 constraint} guarantee that the achievable data rate of each user exceeds the target rate $ R $; the constraint \eqref{eq:phase constraint} guarantees phase alignment by restricting the energy of the residual imaginary component to be below a small tolerance $\epsilon > 0$, thereby ensuring the validity of the real-valued approximation utilized in the linear observation model. The coupled variables of transmit power $ P_{k,t} $ and phase alignment $ \phi $ leads to the non-convextity of problem \eqref{eq:Oriproblem}. To tackle this non-convex issue, in the following, we propose a hierarchical three-layer decomposition framework to address \eqref{eq:Oriproblem}, decoupling it into three tractable sub-problems, resolved in a sequential and iterative manner. 

\section {Proposed Three-Layer Decomposition Framework}\label{section: framework}
To resolve the computational intractability caused by the non-convexity and strong variable coupling among optimization variablesidentified in the problem formulation, we propose a hierarchical three-layer decomposition framework, which is described as follows: 
\begin{itemize}
	\item \textbf{Inner Layer: Phase Alignment and Channel Estimation.} 
    Serving as the physical layer foundation, this module tackles the hardware-induced phase ambiguity. Specifically, for a given power allocation set, a tailored EM-GS algorithm is developed to jointly estimate the user symbols and optimize the receiver's phase compensation vector $\boldsymbol{\psi}$. Consequently, the derived effective real-valued channel gain, $G_{k,t}$, satisfies the linearization constraint in \eqref{eq:simplified form of the signal}, thereby furnishing accurate channel state information (CSI) for the upper layers.
	
	\item \textbf{Middle Layer: Power Allocation.}
	This layer serves as the core resource management component. By utilizing the phase-calibrated, real-valued effective channel gains provided by the Inner Layer, the originally non-convex power allocation problem is transformed into a convex formulation. We solve this efficiently using a Lagrangian multiplier method derived from the KKT conditions. This yields a closed-form solution that dynamically allocates transmit power $P_{k,t}$ to minimize total system consumption while strictly satisfying the target rate requirements $R$.

    Serving as the core for resource management, this layer leverages the phase-calibrated effective channel gains derived from the Inner Layer. Consequently, the originally non-convex power allocation problem is reformulated as a convex optimization problem. By analyzing the KKT conditions, a closed-form solution is derived to efficiently solve the problem. This solution dynamically allocates the transmit power $P_{k,t}$ to minimize total power consumption, while strictly satisfying the target rate requirements $R$.

	\item \textbf{Outer Layer: Online Signal Detection. }
	This layer is responsible for the real-time system operation after the Inner and Middle layers have optimized phase alignment and power allocation. It uses a simplified linear channel model and employs low-complexity linear MMSE detection to reliably recover multi-user signals, without resorting to complex non-linear decoding.
\end{itemize}

\subsection{Inner-Layer: Phase Alignment and Channel Estimation}
As delineated in the system model, the RAQR is constrained to acquire amplitude-only observations, denoted as $\mathbf{z}_t$, thereby introducing inherent phase ambiguity. Consequently, the primary objective of the Inner Layer is to mitigate this hardware-induced limitation and furnish the precise, effective real-valued CSI necessitating for the subsequent power allocation layer. Specifically, given a fixed transmit power allocation set $\{P_{k,t}\}$, a customized EM-GS algorithm is used to jointly estimate the user transmitted signals, $s_{k,t}$, and optimize the receiver's phase compensation vector, $\boldsymbol{\psi}$.

Based on the observation model in \eqref{eq:Observation model}, let $\mathbf{s}'_t \triangleq [s_1 e^{j\psi_1}, s_2 e^{j\psi_2}]^T$ denote the equivalent phase-shifted symbol vector, and $\mathbf{A}_t \triangleq [\mathbf{h}_{1,t}, \mathbf{h}_{2,t}]$ represent the aggregated channel matrix. Consequently, the observation equation can be recast in a compact vector form as follows:
\begin{align}
	\mathbf{z}_t = \left| \mathbf{A}_t \mathbf{s}'_t + \mathbf{b} + \mathbf{n}_t \right|.
	\label{eq:Equivalent observation model}
\end{align}

It is noteworthy that this formulation is mathematically isomorphic to the biased phase retrieval model established in \cite{cui2025towards}. Consequently, symbol detection in our setting can be regarded as a phase-retrieval problem, where the unknown symbol phases are treated as latent variables and the complex vector $\mathbf{s}'_t$ is reconstructed solely from magnitude-based measurements. 

To address the non-convexity inherent in this problem, the EM-GS algorithm introduces a set of latent variables defined as the unobserved phase vector of the received signal at the antenna array, denoted by $\boldsymbol{\theta}_t = [\theta_{1,t}, \ldots, \theta_{M,t}]^T \in \mathbb{R}^{M}$. The underlying rationale is that, provided the received phases $\boldsymbol{\theta}_t$ are known, the complex signal recovery task reduces to a computationally tractable linear Gaussian estimation problem.

The algorithm proceeds by alternating between two steps: (1) the Expectation (E)-step, which computes the posterior statistics of the latent phase vector $\boldsymbol{\theta}_t$ conditioned on the current parameter estimates; and (2) the Maximization (M)-step, which updates $\mathbf{s}'_t$ (and by extension $\boldsymbol{\psi}$) to maximize the expected log-likelihood function. This iterative procedure effectively retrieves the missing phase information, thereby facilitating the linearization of the effective channel model.

\subsubsection{E-step: Phase posterior expectation calculation} 
In the $i$-th iteration, conditioned on the current symbol estimate $\mathbf{s}^{\prime}(i)$, the conditional probability density function of the observation $z_{m,t}$ is characterized by a Rician distribution:
	\begin{align}
		& p(z_{m,t}|\mathbf{s}^{\prime{(i)}})\nonumber\\ &~=\frac{2z_{m,t}}{\sigma^2}\exp\left(-\frac{z_{m,t}^2+|\lambda_{m,t}^{(i)}|^2}{\sigma^2}\right) I_0\left(\frac{2z_{m,t}|\lambda_{m,t}^{(i)}|}{\sigma^2}\right),
		\label{eq:Rician}
	\end{align}
where $\lambda_{m,t}^{(i)}=\mathbf{a}_{m,t}^T\mathbf{s}^{\prime(i)}+b_m\mathrm{,}\mathbf{a}_{m,t}^T$, $\mathbf{a}_{m,t}^T$ is the $m$-th row of $\mathbf{A}$ in time slot $t$, $I_0(\cdot)$ is the modified Bessel function of the first kind of order zero.

Accordingly, the posterior distribution of the latent phase $\theta_{m,t}$ follows a von Mises distribution, given by:
\begin{align}
	p(\theta_{m,t}|z_{m,t},\mathbf{s}^{\prime{(i)}})=\frac{\exp\left(\kappa_{m,t}^{(i)}\cos(\theta_{m,t}-\vartheta_{m,t}^{(i)})\right)}{2\pi I_0(\kappa_{m,t}^{(i)})},
		\label{eq:von Mises}
\end{align}
where $\vartheta_{m,t}^{(i)}=\angle\lambda_{m,t}^{(i)}$ represents the mean direction (or phase center) under the current estimate. The term $\kappa_{m,t}^{(i)}=\frac{2z_{m,t}|\lambda_{m,t}^{(i)}|}{\sigma^2}$ denotes the concentration parameter, which quantifies the reliability of the phase estimation, while $\lambda_{m,t}^{(i)}=\mathbf{a}_{m,t}^{T}\mathbf{s}^{\prime(i)}+b_{m}$ corresponds to the estimated noiseless complex signal.

\subsubsection{M-step: Update of Symbol Estimates}
Subsequently, the symbol estimates are updated by maximizing the expected complete-data log-likelihood function. Solving this optimization problem yields the following closed-form expression:
\begin{align}
		\mathbf{s}^{\prime(i+1)}=(\mathbf{A}_t^T\mathbf{A}_t)^{-1}\mathbf{A}_t^T\left(\mathbf{z}_t\circ e^{j\boldsymbol{\vartheta}_t^{(i)}}\circ R(\boldsymbol{\kappa}_t^{(i)})-\mathbf{b}\right),
		\label{eq:The closed-form solution}
\end{align}
where $R(x)=I_1(x)/I_0(x)$ acts as a Signal-to-Noise Ratio (SNR)-dependent high-pass filter, and the operator $\circ$ denotes the element-wise Hadamard product.

\subsubsection{Gradient derivation of phase updates}
With the objective of directly optimizing the phase compensation vector $\boldsymbol{\psi}$ to eliminate channel phase distortion, an explicit gradient descent-based update step is incorporated subsequent to each M-step of the EM-GS algorithm. To this end, the phase alignment cost function is formulated as the aggregate power of the imaginary components of the equivalent channel:
\begin{align}
		J(\boldsymbol{\psi})=\sum_{m=1}^M\sum_{k=1}^2\left|\Im\{h_{m,k}e^{j\psi_k}\}\right|^2.
		\label{eq:The phase alignment cost function}
\end{align}

Minimizing the objective function in \eqref{eq:The phase alignment cost function} inherently drives the equivalent channel coefficients $h_{m,k}e^{j\psi_k}$ towards the real domain, thereby effectively mitigating phase ambiguity. To facilitate the subsequent gradient derivation, we decompose the channel gain as $h_{m,k}=a_{m,k}+jb_{m,k}$. By applying Euler's formula, the real and imaginary components of the phase-compensated channel can be explicitly defined as:
\begin{align}
		\Re\{h_{m,k}e^{j\psi_{k}}\} &= a_{m,k}\cos\psi_{k}-b_{m,k}\sin\psi_{k} \triangleq R_{m,k}, \\
		\Im\{h_{m,k}e^{j\psi_{k}}\} &= a_{m,k}\sin\psi_{k}+b_{m,k}\cos\psi_{k} \triangleq I_{m,k},
    \label{eq:Real_Imag_Components}
\end{align}
where $R_{m,k}$ and $I_{m,k}$ denote the real and imaginary terms, respectively.

Leveraging the definitions in \eqref{eq:Real_Imag_Components}, the gradient of the cost function can be derived via the chain rule. A critical observation is that the derivative of the imaginary component with respect to the phase parameter yields the real component, i.e., $\frac{\partial I_{m,k}}{\partial\psi_k} = R_{m,k}$. Consequently, the partial derivative of the quadratic term simplifies to:
\begin{align}
		\frac{\partial\left(I_{m,k}^2\right)}{\partial\psi_k} &= 2 I_{m,k} \cdot \frac{\partial I_{m,k}}{\partial\psi_k} \nonumber\\
        &= 2 \cdot \underbrace{\Im\{h_{m,k}e^{j\psi_k}\}}_{I_{m,k}} \cdot \underbrace{\Re\{h_{m,k}e^{j\psi_k}\}}_{R_{m,k}}  
         = 2 I_{m,k} R_{m,k}.\nonumber
\end{align}

Aggregating the partial derivatives for each user $k \in \{1, 2\}$, the gradient of the cost function $J(\boldsymbol{\psi})$ can be compactly expressed in vector form as:
\begin{align}
    \nabla J(\boldsymbol{\psi}) & =
    \left[ \frac{\partial J}{\partial\psi_1}, \frac{\partial J}{\partial\psi_2} \right]^T \nonumber\\& = 2
    \begin{bmatrix}
        \sum_{m=1}^M \Im\{h_{m,1}e^{j\psi_1}\} \Re\{h_{m,1}e^{j\psi_1}\} \\
        \sum_{m=1}^M \Im\{h_{m,2}e^{j\psi_2}\} \Re\{h_{m,2}e^{j\psi_2}\}
    \end{bmatrix}.
    \label{eq:gradient_vector}
\end{align}
The availability of this analytical gradient facilitates an efficient update using the gradient descent method. The iterative update rule is given by:
\begin{align}
    \boldsymbol{\psi}^{(i+1)} = \boldsymbol{\psi}^{(i)} - \mu \nabla J(\boldsymbol{\psi}^{(i)}),
    \label{eq:gradient_update}
\end{align}
where $i$ denotes the iteration index, and $\mu > 0$ represents the step size (or learning rate), which governs the convergence speed.

In summary, the proposed inner-layer optimization framework achieves joint estimation of the user symbols $\mathbf{s}_t$ and phase parameters $\boldsymbol{\psi}$ by orchestrating the EM-GS algorithm with the gradient descent updates. The resulting optimal phase vector $\boldsymbol{\psi}^{*}$ and the corresponding equivalent real channel gains, calculated as $G_{k,t}=\|\Re\{\mathbf{h}_{k,t}\circ e^{j\boldsymbol{\psi}^*}\}\|^2$, are then fed as precise CSI into the subsequent outer-layer power allocation module.
\subsection{Middle-Layer: Power Allocation}\label{subsec:Power Allocation}
Once the inner layer converges, the system obtains the optimal phase compensation $\psi^{*}$ and the corresponding equivalent real channel gain $G_{k,t}=\|\Re\{\mathbf{h}_{k,t}\circ e^{j\psi^{*}}\}\|^{2}$. At this point, we need to determine whether the outer layer power allocation problem is convex. In the convexity proof, we will write the optimization problem as follows:
\begin{subequations}\label{eq:outer_opt_problem}
		\begin{align}
			\min_{\substack{P_{1,1},P_{1,2},P_{2,2}}} \quad & P_{\text{total}} = P_{1,1} + P_{1,2} + P_{2,2} \label{eq:obj} \\
			\text{s.t.} \quad & R^{\text{target}} - \left[ \log_{2}\left(1 + \frac{P_{1,1}G_{1,1}}{\sigma^{2}}\right) \right. \nonumber\\ &~ \left. + \log_{2}\left(1 + \frac{P_{1,2}G_{1,2}}{P_{2,2}G_{2,2} + \sigma^{2}}\right) \right] \leq 0, \label{eq:con_rate1} \\
			& R^{\text{target}} - \log_{2}\left(1 + \frac{P_{2,2}G_{2,2}}{\sigma^{2}}\right) \leq 0, \label{eq:con_rate2} \\
			& -P_{1,1} \leq 0,\; -P_{1,2} \leq 0,\; -P_{2,2} \leq 0, \label{eq:con_nonneg} \\
			& P_{1,1} - P_{1,\max} \leq 0,\; P_{1,2} - P_{1,\max} \leq 0,\; \nonumber\\ &~ P_{2,2} - P_{2,\max} \leq 0. \label{eq:con_maxpower}
		\end{align}
\end{subequations}

The feasibility of the optimization problem is subject to the non-negativity and maximum power limits specified in \eqref{eq:con_nonneg} and \eqref{eq:con_maxpower}. These linear inequalities define a convex polyhedral set. Furthermore, since the objective function in \eqref{eq:obj} and the constraints in \eqref{eq:con_nonneg}--\eqref{eq:con_maxpower} are affine, establishing the convexity of the problem reduces to analyzing the rate constraints.

Consider the constraint function for user 2 in \eqref{eq:con_rate2}, denoted as:
\begin{align}
		f_{C2}(\mathbf{P}) = R^{\text{target}} - \log_2\left(1+\frac{P_{2,2}G_{2,2}}{\sigma^2}\right).
		\label{eq:f_{C2}}
\end{align}
Let $g(P) = -\log_2(1+aP)$ with $a = G_{2,2}/\sigma^{2} > 0$. The convexity of $f_{C2}$ hinges on the convexity of $g(P)$. By examining the second-order derivative:
\begin{align}
		g^{\prime\prime}(P) = \frac{a^2}{\ln(2)(1+aP)^2} > 0,
		\label{eq:convexity_proof}
\end{align}
we confirm that $g(P)$ is strictly convex for $P \ge 0$, thereby proving the convexity of constraint \eqref{eq:con_rate2}.\\
\indent However, the constraint function $f_{C1}$ associated with \eqref{eq:con_rate1} is shown on top of the next page that expressed as:
\begin{figure*}[!t]
	\begin{align}
		f_{C1}(\mathbf{P}) = R^{\text{target}} - \left[ \underbrace{\log_2\left(1+\frac{P_{1,1}G_{1,1}}{\sigma^2}\right)}_{h_1(P_{1,1})} + \underbrace{\log_2\left(1+\frac{P_{1,2}G_{1,2}}{P_{2,2}G_{2,2}+\sigma^2}\right)}_{h_2(P_{1,2},P_{2,2})} \right],
		\label{eq:f_{C1}}
	\end{align}
    \hrulefill
\end{figure*}
is non-convex due to the coupling of variables $P_{1,2}$ and $P_{2,2}$ within the term $h_2$. Consequently, the overall optimization problem is generally non-convex. To address this, we exploit the monotonicity of the objective function and the rate constraints. Since minimizing the total power requires the rate constraints to be active (i.e., satisfied with equality), the optimal solution must lie on the boundary of the feasible region. By converting the inequality constraint \eqref{eq:con_rate2} into an equality, we can directly derive the optimal power allocation for user 2, $P_{2,2}^*$, as:
\begin{align}	
		 \! \log_2\left(1\!+\!\frac{P_{2,2}G_{2,2}}{\sigma^{2}}\right) \!=\! R^{\text{target}}  
		\Rightarrow    P_{2,2}^{*}\!=\!\frac{\sigma^2(2^{R^{\text{target}}}\!-\!1)}{G_{2,2}}.\!\!\!\!\label{eq:p22_opt}
\end{align}

Substituting the optimal $P_{2,2}^*$ into the original formulation reduces the problem to determining the optimal power distribution for User 1. The rate constraint $f_{C1}$ decomposes into a sum of two univariate logarithmic functions with respect to $P_{1,1}$ and $P_{1,2}$. Since the sum of concave functions preserves concavity, the feasible region defined by the rate requirement is convex. Consequently, minimizing the affine objective function (total power) over this convex set constitutes a standard convex optimization problem.

Assuming the feasible region is non-empty (satisfying Slater's condition), the KKT conditions provide necessary and sufficient criteria for optimality. The rate constraint \eqref{eq:f_{C1}} is rewritten as:
\begin{align}
		\!\!\!  \log_2\left(1+\frac{P_{1,1}G_{1,1}}{\sigma^2}\right)+\log_2\left(1+\frac{P_{1,2}G_{1,2}}{I}\right)\geq R^{\text{target}},\nonumber
\end{align}
where $I=P_{2,2}^*G_{2,2}+\sigma^2$ represents the constant aggregate noise-plus-interference term. The Lagrangian function is formulated as:
\begin{align}
	\begin{split}
		\mathcal{L}(P_{1,1}, P_{1,2}, \lambda) &= P_{1,1} + P_{1,2} - \lambda \bigg[ \log_2 \left( 1 + \frac{P_{1,1}G_{1,1}}{\sigma^2} \right) \\
		&\quad + \log_2 \left( 1 + \frac{P_{1,2}G_{1,2}}{I} \right) - R^{\text{target}} \bigg].
	\end{split}
\end{align} 
Applying the stationarity condition of the KKT system, we set the partial derivatives of the Lagrangian with respect to $P_{1,1}$ and $P_{1,2}$ to zero, yielding the following closed-form expressions:
\begin{align}
		\!\!\! P_{1,1}=\frac{\lambda}{\ln 2}-\frac{\sigma^2}{G_{1,1}},~
		P_{1,2}=\frac{\lambda}{\ln 2}-\frac{I}{G_{1,2}}. \!\!\!\!
		\label{eq:stationarity_condition}
\end{align}
\indent Invoking the complementary slackness condition from the KKT framework, we observe that the optimal solution typically requires the rate constraint to be active, implying that the inequality holds with equality. Consequently, we have:
\begin{align}
		\log_2\left(1+\frac{P_{1,1}G_{1,1}}{\sigma^2}\right)+\log_2\left(1+\frac{P_{1,2}G_{1,2}}{I}\right)=R^{\text{target}}. \nonumber
\end{align}
Substituting the stationarity expressions from \eqref{eq:stationarity_condition} into this equality allows us to solve for the optimal Lagrange multiplier $\lambda^{*}$:
\begin{align}
		\lambda^*=\ln 2\cdot 2^{R^{\text{target}}/2}\cdot\sqrt{\frac{\sigma^2I}{G_{1,1}G_{1,2}}}.
		\label{eq:lambda}
\end{align}
Recalling from the optimization of User 2 that $I = P_{2,2}^*G_{2,2}+\sigma^2 = \sigma^2 2^{R^{\text{target}}}$, we substitute this relationship and $\lambda^*$ back into \eqref{eq:stationarity_condition}. This yields the final closed-form power allocation for User 1:
\begin{subequations}\label{eq:power_closed_form}
    \begin{align}
	P_{1,1}^{*}&=\sigma^{2}\left(\frac{2^{R^{\text{target}}}}{\sqrt{G_{1,1}G_{1,2}}}-\frac{1}{G_{1,1}}\right),\\
		P_{1,2}^*&=\sigma^2\left(\frac{2^{R^{\text{target}}}}{\sqrt{G_{1,1}G_{1,2}}}-\frac{2^{R^{\text{target}}}}{G_{1,2}}\right).
\end{align}
\end{subequations}
\subsection{Outer-Layer: Online Signal Detection}
Upon the convergence of the alternating optimization between the inner and middle layers, the system establishes a steady-state operating point characterized by the optimal phase compensation vector $\boldsymbol{\psi}^* = [\psi_1^*, \psi_2^*]^T$ and the optimal power allocation set $\{P_{1,1}^*, P_{1,2}^*, P_{2,2}^*\}$. With these parameters fixed, the objective of the third layer is to execute high-throughput, real-time detection of the transmitted data payload. In contrast to the iterative estimation required in the Inner Layer, this stage requires a low-complexity, non-iterative solution suitable for continuous data streams.

\subsubsection{Phase Pre-compensation}
Utilizing the optimized phase vector $\boldsymbol{\psi}^*$, phase compensation is applied to the complex baseband signal $\mathbf{y}_t$ received in each time slot $t$. This operation aligns the channel phase with that of the LO signal $\mathbf{b}$, yielding the rotated signal:
\begin{align}
		\tilde{\mathbf{y}}_t=\mathbf{y}_t\circ e^{-j\boldsymbol{\psi}^*}.
\end{align} 

\subsubsection{Signal Linearization and Detection}
Following phase alignment, the signal structure simplifies to the form presented in \eqref{eq:simplified form of the signal}. By subtracting the known amplitude of the LO signal, we establish an approximate linear real-valued observation model:
\begin{align}
		\hat{\mathbf{z}}_{t} = |\tilde{\mathbf{y}}_{t}| - |\mathbf{b}| \approx \mathbf{H}_{t}^{\mathrm{opt}}\mathbf{s}_{k,t} + \tilde{\mathbf{n}}_{t},
    \label{eq:linear_model}
\end{align} 
where $\mathbf{H}_t^{\mathrm{opt}}=\Re\{\mathbf{H}_t\circ e^{j\boldsymbol{\psi}^*}\}\in\mathbb{R}^{M\times2}$ denotes the known optimal real-valued equivalent channel matrix.

\subsubsection{Linear MMSE Detection}
Leveraging the linear real-valued observation model established in \eqref{eq:linear_model}, standard linear estimation techniques becomes applicable. We employ a Linear MMSE detector, chosen for its robustness in interference-limited regimes. The LMMSE detector minimizes the mean square error between the transmitted symbols and their estimates, effectively balancing Gaussian noise suppression and the mitigation of residual inter-user interference inherent in the Hybrid NOMA protocol. The estimated symbol vector is given by:
\begin{align}
		\begin{bmatrix}
			\hat{s}_{1,t} \\
			\hat{s}_{2,t}
\end{bmatrix}_{\mathrm{MMSE}}=\left((\mathbf{H}_t^\mathrm{opt})^T\mathbf{H}_t^\mathrm{opt}+\sigma^2\mathbf{I}_2\right)^{-1}(\mathbf{H}_t^\mathrm{opt})^T\hat{\mathbf{z}}_t.
\end{align} 

\subsubsection{Hard Decision}
The final demodulation stage involves mapping the continuous soft estimates $\hat{s}_{k,t}$ obtained from the MMSE detector to the nearest constellation points in the modulation scheme. This is achieved via a euclidean distance minimization, mathematically represented as:
\begin{align}
		\bar{s}_{k,t}=\mathcal{Q}(\hat{s}_{k,t}),
\end{align} 
where $\mathcal{Q}(\cdot)$ represents the PAM demapping operation.

Here is the complete process of the three-layer alternating optimization framework, where $\varepsilon > 0$ is a very small value to denote the convergence threshold. This three-layer alternating optimization algorithm can be summarized in \emph{Algorithm} \ref{alg:alternating_optimization}.

\vspace{0.5em}
\hrule 
\vspace{0.5em}

\begin{algorithm}\label{alg:alternating_optimization}
	Three-layer alternating optimization for atomic-receiver-assisted Hybrid NOMA
	\vspace{0.5em}
	\hrule
	\vspace{0.5em}
	\begin{enumerate}
		\item \textbf{Given}: Initial power allocation $\mathbf{P}^{(0)} = \{P_{k,t}^{(0)}\}$, phase vector $\bm{\psi}^{(0)}$, tolerance $\delta > 0$, maximum outer iterations $N_{\max}$.
		\begin{enumerate}
			\item Set $n \gets 0$, $P_{\text{total}}^{(0)} \gets \sum_{k,t} P_{k,t}^{(0)}$.
			\item \textbf{Repeat}: optimize $\bm{\psi}$ and $\mathbf{P}$ in an alternating manner.
			\begin{enumerate}
				\item \textbf{Inner-Layer: Phase Alignment and Channel Estimation}
				\begin{enumerate}
					\item[\textbf{Step 1:}] Fix $\mathbf{P}^{(n)}$, we find the optimal phase vector $\bm{\psi}^{(n+1)}$ by using the EM-GS algorithm until convergence.
					\item[\textbf{Step 2:}] Compute equivalent channel gains: $G_{k,t}^{(n+1)} = \|\Re\{\mathbf{h}_{k,t} \circ e^{j\bm{\psi}^{(n+1)}}\}\|^2$.
				\end{enumerate}
				
				\item \textbf{Middle-Layer: Power Allocation}
				\begin{enumerate}
					\item[\textbf{Step 1:}] Fix $G_{k,t}^{(n+1)}$, we find the optimal power allocation $\mathbf{P}^{(n+1)}$ by solving the convex problem via KKT conditions.
					\item[\textbf{Step 2:}] Compute total power: $P_{\text{total}}^{(n+1)} = \sum_{k,t} P_{k,t}^{(n+1)}$.
				\end{enumerate}
				
				\item \textbf{Convergence Check}
				\begin{enumerate}
					\item[\textbf{Step 1:}] Compute relative change: $\Delta = |P_{\text{total}}^{(n+1)} - P_{\text{total}}^{(n)}| / P_{\text{total}}^{(n)}$.
					\item[\textbf{Step 2:}] $n \gets n + 1$.
				\end{enumerate}
			\end{enumerate}
			\item \textbf{Until} $\Delta < \delta$ or $n \geq N_{\max}$.
		\end{enumerate}
		
		\item \textbf{Outer-Layer: Online Signal Detection}
		\begin{enumerate}
			\item \textbf{Phase Pre-compensation}: Apply optimal phase $\bm{\psi}^* = \bm{\psi}^{(n)}$ to received signal $\mathbf{y}_t$: $\tilde{\mathbf{y}}_t = \mathbf{y}_t \circ e^{-j\bm{\psi}^*}$.
			\item \textbf{Amplitude Extraction}: Obtain linear real observation: $\hat{\mathbf{z}}_t = |\tilde{\mathbf{y}}_t| - |\mathbf{b}|$.
			\item \textbf{Linear MMSE Detection}: Compute symbol estimates:
			\[
			\hat{\mathbf{s}}_t = \big( (\mathbf{H}_t^{\mathrm{opt}})^T \mathbf{H}_t^{\mathrm{opt}} + \sigma^2 \mathbf{I}_2 \big)^{-1} (\mathbf{H}_t^{\mathrm{opt}})^T \hat{\mathbf{z}}_t,
			\]
			where $\mathbf{H}_t^{\mathrm{opt}} = \Re\{\mathbf{H}_t \circ e^{j\bm{\psi}^*}\}$.
			\item \textbf{Symbol Decision}: Demodulate via PAM: $\bar{s}_{k,t} = \mathcal{Q}(\hat{s}_{k,t}), \quad \forall k,t$.
		\end{enumerate}
		
		\item \textbf{Output}: 
		\begin{itemize}
			\item Optimal phase: $\bm{\psi}^* = \bm{\psi}^{(n)}$
			\item Optimal power: $\mathbf{P}^* = \mathbf{P}^{(n)}$
			\item Equivalent channels: $G_{k,t}^*$
			\item Detected symbols: $\{\bar{s}_{k,t}\}$ for all $k,t$
		\end{itemize}
	\end{enumerate}
	\vspace{0.5em}
	\hrule
	\vspace{0.5em}
\end{algorithm}

\section{Complexity and Convergence Analysis}
\label{sec:complexity_convergence}
To validate the practical feasibility and theoretical reliability of the proposed three-layer framework, we present a rigorous analysis of its computational complexity and convergence behavior.

\subsubsection{Computational Complexity Analysis}
\label{subsec:complexity}
The computational burden is distributed across the three hierarchical layers. We analyze the complexity of each layer and the overall system as follows:

\begin{itemize}
	\item \textbf{Inner Layer (Phase Alignment and Channel Estimation):} 
	The computational cost here is dominated by the Expectation-Step in the EM-GS algorithm and the gradient calculation for phase updates. Given the system configuration, these operations involve matrix--vector multiplications that scale linearly with the number of atomic antennas $M$. Specifically, for a fixed number of users, the complexity per inner iteration is $\mathcal{O}(M)$. Assuming the inner layer requires $I_{\text{in}}$ iterations to converge, the cumulative complexity is $\mathcal{O}(I_{\text{in}} \cdot M)$.
	
	\item \textbf{Middle Layer (Power Allocation):} 
	This layer leverages the derived analytical closed-form solutions (as seen in \eqref{eq:p22_opt} and \eqref{eq:power_closed_form}). Since these computations involve only scalar arithmetic operations and elementary functions (logarithms/exponentials), the complexity is independent of the array size $M$, i.e., $\mathcal{O}(1)$. Consequently, the computational overhead of the middle layer is negligible compared to the inner layer.
	
	\item \textbf{Outer Layer (Online Signal Detection):} 
	Once the parameters are optimized, the outer layer performs phase pre-compensation and linear MMSE detection. The pre-compensation involves element-wise vector operations with complexity $\mathcal{O}(M)$. The MMSE detection can be written as a multiplication between a $1 \times M$ detection row vector and an $M \times M$ matrix, yielding a $1 \times M$ output vector. Each of the $M$ output entries is a dot product of length $M$, resulting in $M \cdot M = M^2$ multiplications and $M \cdot (M-1) = M^2 - M$ additions. Therefore, the overall online detection complexity is dominated by this operation and scales quadratically with the array size, i.e., $\mathcal{O}(M^2)$.
	
	\item \textbf{Total Optimization Complexity:} 
	The joint optimization executes an alternating process between the Inner and Middle layers. Denoting the number of iterations between these two layers required for convergence as $I_{\text{loop}}$, the total worst-case computational complexity for the optimization phase is $\mathcal{O}(I_{\text{loop}} \cdot I_{\text{in}} \cdot M)$, while the subsequent online detection stage incurs an additional per-symbol complexity of $\mathcal{O}(M^2)$.
\end{itemize}

\textit{Comparison with Benchmarks:} The proposed framework demonstrates a significant efficiency advantage. Its complexity scales linearly with the number of atomic antennas ($\mathcal{O}(M)$). In stark contrast, standard Semi-Definite Programming (SDP) solvers typically exhibit a polynomial complexity of at least $\mathcal{O}(M^{3.5})$, while exhaustive search methods suffer from exponential scaling. This linear scaling characteristic renders the proposed approach highly suitable for large-scale RAQR systems.

\subsubsection{Convergence Analysis}
\label{subsec:convergence}
The optimization process within the framework operates effectively as a Block Coordinate Descent (BCD) method iterating between the Inner and Middle layers. We analyze the convergence properties as follows:

\begin{itemize}
	\item \textbf{Inner Layer Convergence:} The convergence is underpinned by the theoretical properties of the EM algorithm and gradient descent. The EM-GS component follows the monotonic likelihood ascent principle, guaranteeing that the expected complete-data log-likelihood is non-decreasing in each step. Simultaneously, the gradient descent for phase alignment minimizes the residual imaginary energy. With an appropriately chosen step size $\mu$, this generates a bounded, non-increasing sequence of the cost function values. By the Monotone Convergence Theorem, the inner layer is guaranteed to converge to a stationary point.
	
	\item \textbf{Global Convergence (Inner-Middle Interaction):} The alternating optimization iteratively updates two blocks of variables: the phase/channel parameters (Inner Layer) and the power allocation (Middle Layer). The mechanism is as follows: 
    (i) The Inner Layer maximizes the equivalent real channel gains $G_{k,t}$ via optimal phase alignment; 
    (ii) The Middle Layer, which is a convex problem, utilizes these improved gains to satisfy rate constraints with strictly lower (or equal) transmit power. 
    Consequently, the total power objective function serves as a Lyapunov function, forming a monotonically non-increasing sequence $\{P_{\text{total}}^{(n)}\}$. Since the transmit power is physically lower-bounded by zero, the sequence must converge.
    
    \item \textbf{Stability of Outer Layer:} Although the Outer Layer (Detection) does not participate in the optimization loop, its performance relies on the convergence of the first two layers. The guaranteed convergence of the Inner and Middle layers ensures that the Outer Layer receives stable, optimal CSI and power parameters, thereby ensuring reliable symbol detection.
\end{itemize}

\section{Simulation Results and Analysis}
\label{sec:simulation}

In this section, we present numerical results to evaluate the performance of the proposed RAQR-based Hybrid NOMA framework. 

\subsection{Convergence Analysis of the Algorithm}
\begin{figure}[!htbp]
	\centering
    \includegraphics[width=0.85\linewidth]{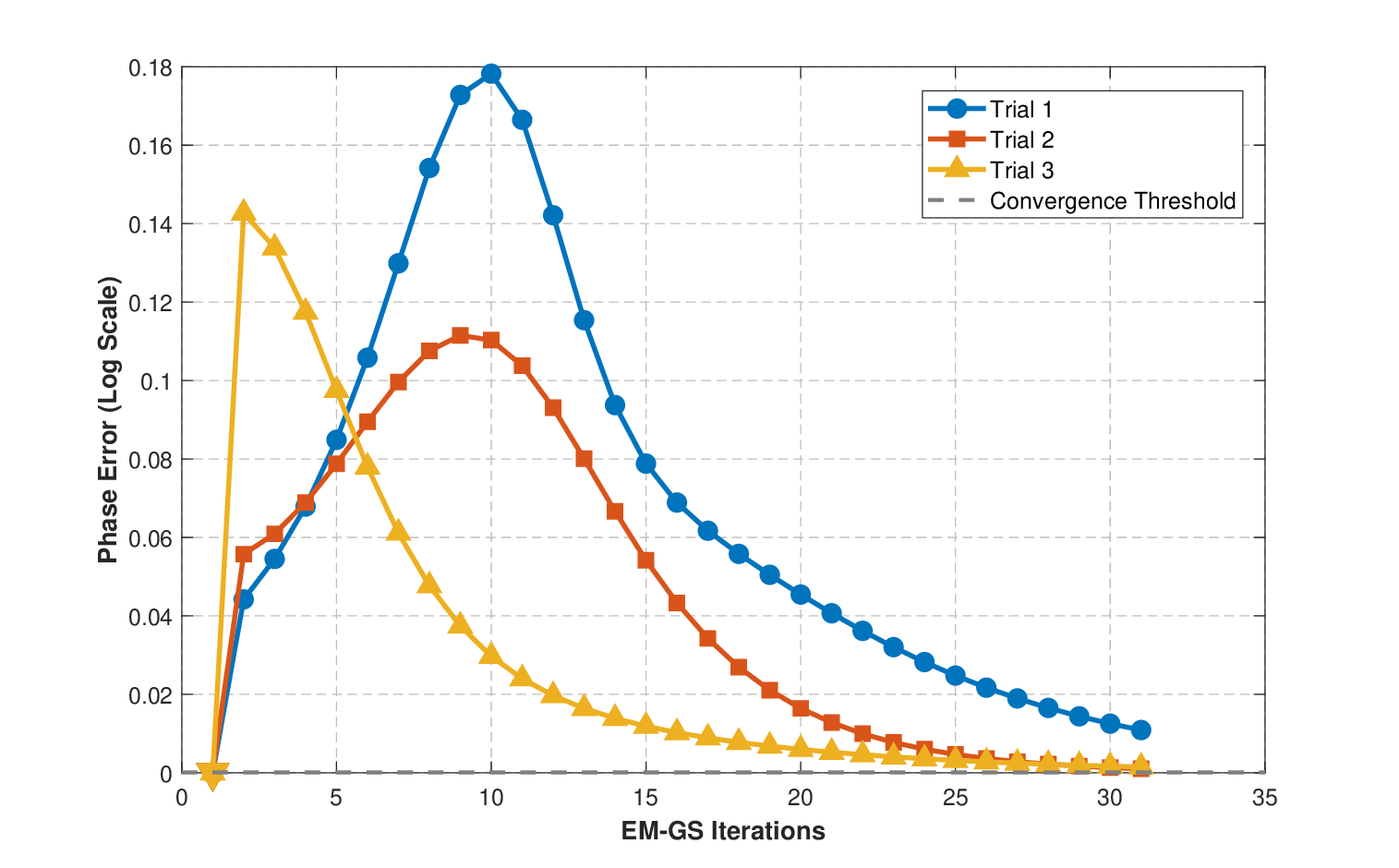}
	\caption{EM-GS: Phase Error Convergence.}
	\label{fig:phase_error}
\end{figure}
As illustrated in Fig. \ref{fig:phase_error}, the phase error evolution presents a two-stage behavior: an initial temporary spike followed by a rapid decay. Notably, the phase error is artificially low (near zero) at the initial iteration ($0$-th iteration). This phenomenon occurs because the algorithm is typically initialized with default reference values before incorporating the actual noisy observations $z_{m,t}$. As the EM-GS algorithm executes its first few iterations, it begins to process the noisy received signals and actively explores the parameter space. This causes the phase estimates to deviate significantly from the naive initialization, resulting in the observed initial spike in the calculated error (peaking between $0.14$ and $0.18$). However, once the gradient search identifies the correct optimal descent direction, the phase error exhibits rapid decay, dropping to approximately $0.06$ -- $0.08$ within the next $10$ to $15$ iterations, before achieving asymptotic stability below $0.02$ after $35$ iterations. This subsequent oscillation-free trajectory aligns with the monotonic convergence theory guaranteed by the EM framework, while the consistent behavior across varying channel initializations (Trials $1$--$3$) confirms the algorithm's robustness. 

\begin{figure}[!htbp]
	\centering
	\includegraphics[width=0.85\linewidth]{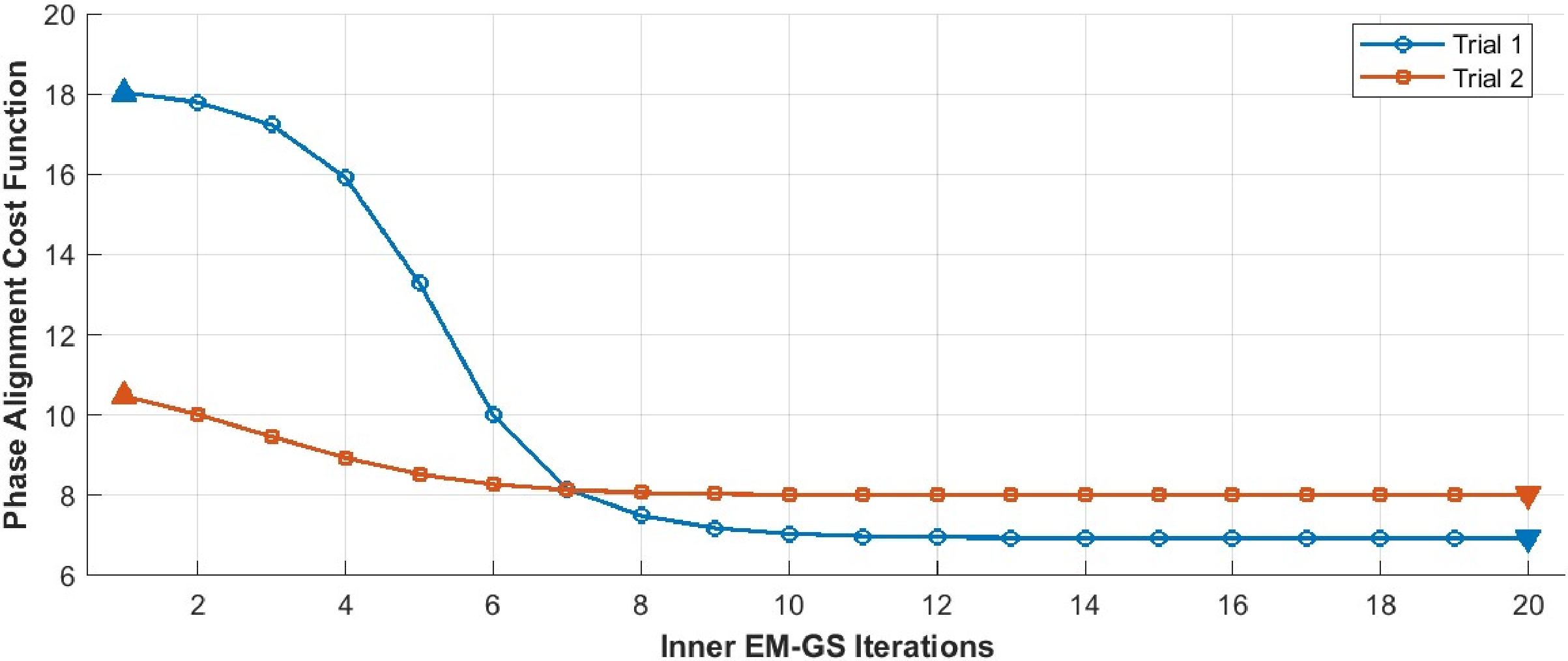}
	\caption{Cost function convergence behavior under different random initializations.}
	\label{fig:convergence}
\end{figure}
The convergence of the inner-layer EM-GS algorithm is depicted in Fig. \ref{fig:convergence}. The residual imaginary energy decreases monotonically, stabilizing after approximately 10 iterations. These consistent trajectories demonstrate insensitivity to initial values, validating the theoretical prediction of effective phase alignment and low-latency interference suppression.

\subsection{Total Power Consumption Analysis}
\begin{figure}[!htbp]
	\centering
	\includegraphics[width=0.85\linewidth]{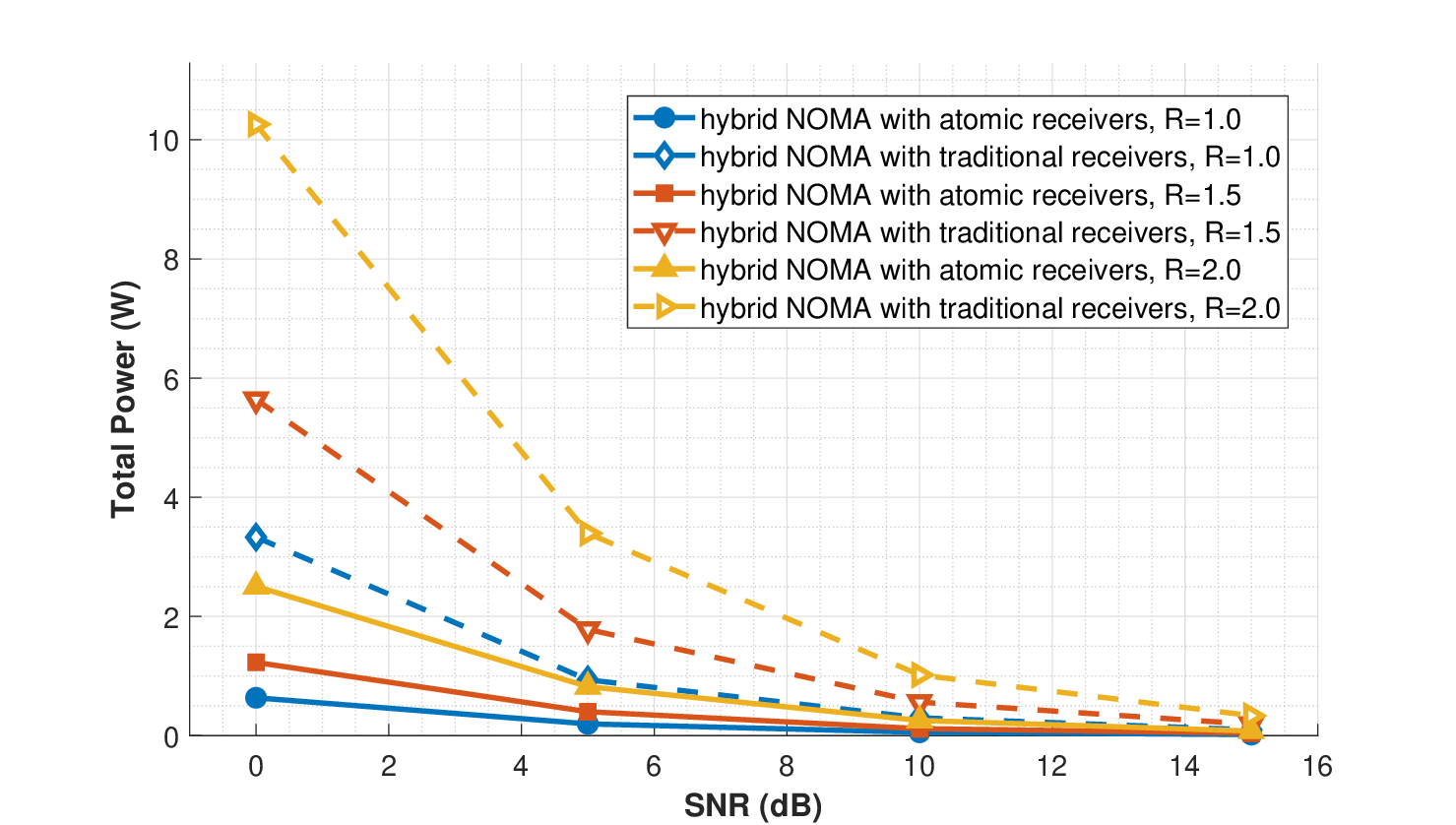}
	\caption{Total Power vs. SNR for Different Data Rates ($M=10$).}
	\label{fig:two transmission_zhexian}
\end{figure}
In Fig. \ref{fig:two transmission_zhexian}, we plot the total transmit power against the reference SNR (i.e., the inherent channel quality determined by path loss and noise variance) for target rates $R \in \{1.0, 1.5, 2.0\}$ bits/s/Hz. As expected, the required power decreases with improving reference SNR and increases with higher $R$. The proposed RAQR-Hybrid NOMA consistently outperforms the traditional benchmark, and the power saving gap widens as $R$ grows, highlighting its scalability for high-throughput demands. This advantage arises from the RAQR's ability to coherently combine signals from multiple vapor cells, maximizing the effective channel gain and thus reducing transmit power. Although atomic receivers can ideally offer up to $20~\textrm{dB}$ sensitivity improvement, the observed gap in this figure is approximately $5-8~\textrm{dB}$ due to the limited number of cells ($M=10$) and multi-user interference. As shown later, scaling up $M$ unleashes the full potential of the atomic architecture.
\subsection{Impact of Atomic Array Size ($M$)}
\begin{figure}[!htbp]
	\centering
	\includegraphics[width=0.85\linewidth]{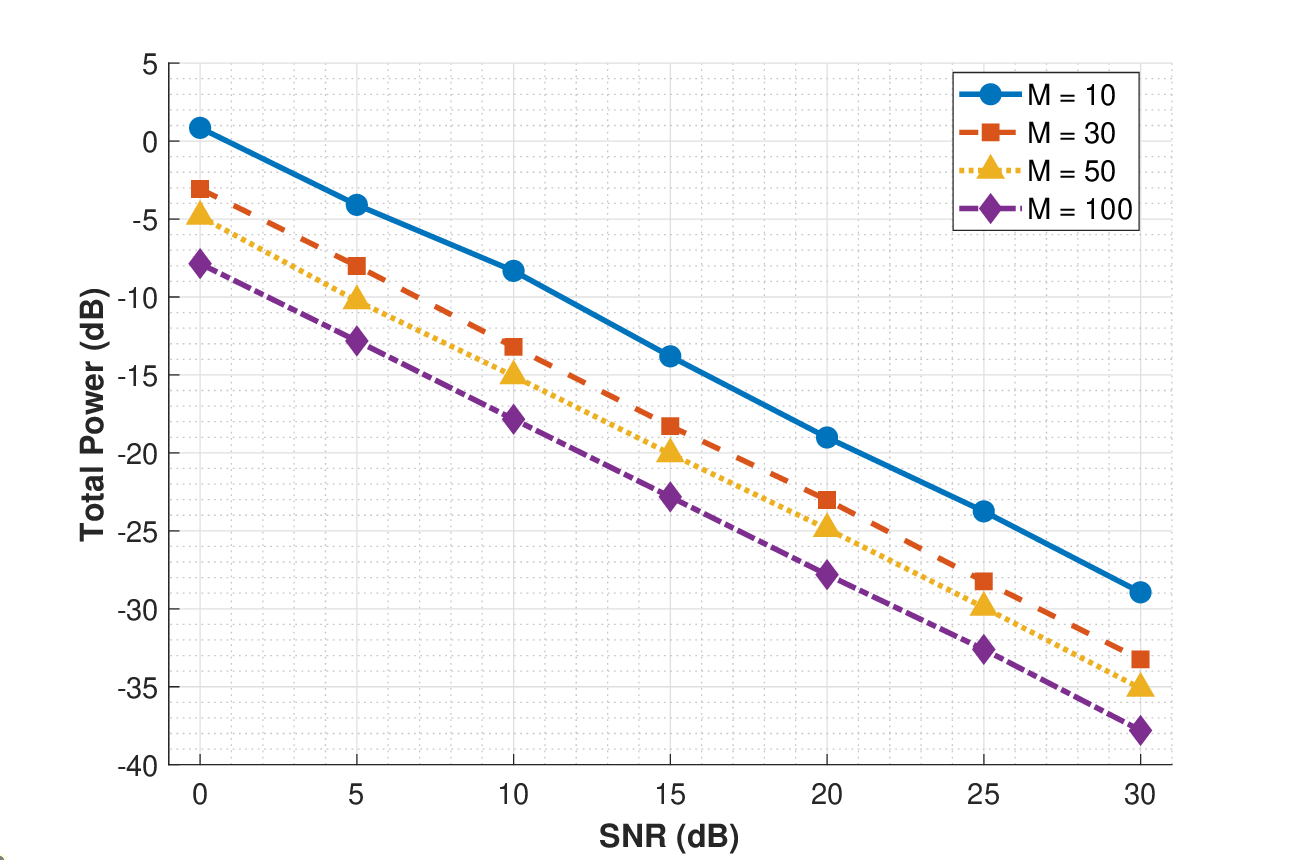}
	\caption{Atomic Receiver Hybrid NOMA: Total Power vs. SNR for Different $M$ Values (Target Rate $R=1.5$ bits/s/Hz).}
	\label{fig:Mzhexian}
\end{figure}
Figure~\ref{fig:Mzhexian} investigates the impact of vapor cell count $M$ on total transmit power for $M \in {10,30,50,100}$ across varying reference SNR levels. Increasing $M$ consistently reduces power consumption, with the most significant drop from $M=10$ to $30$. Beyond that, further increases (e.g., $50\to100$) yield diminishing returns. This behavior stems not only from conventional diversity gain but also from enhanced spatial degrees of freedom that improve intra-cluster interference suppression in the multi-user NOMA setup. Although $M=100$ is currently impractical due to cost and complexity, this exploration reveals the asymptotic potential; a moderate array ($M\approx30$--$50$) can offer a favorable trade-off between energy efficiency, interference mitigation, and hardware feasibility.
\begin{figure}[!htbp]
	\centering
	\includegraphics[width=0.85\linewidth]{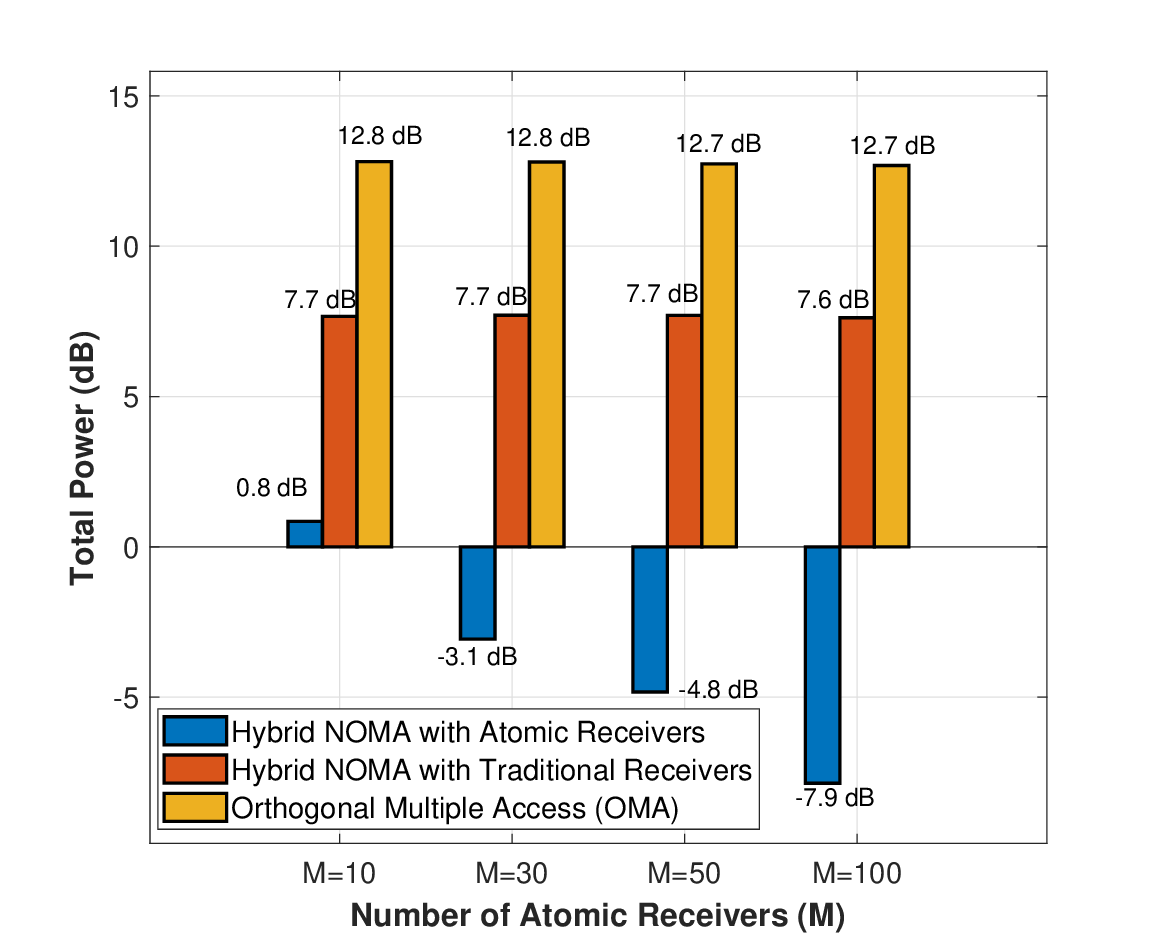}
	\par\vspace{2mm}
	\centering\textbf{(a)} SNR = 0 dB
	\label{fig:0dB}
	\vspace{0.5cm} 
	\includegraphics[width=0.85\linewidth]{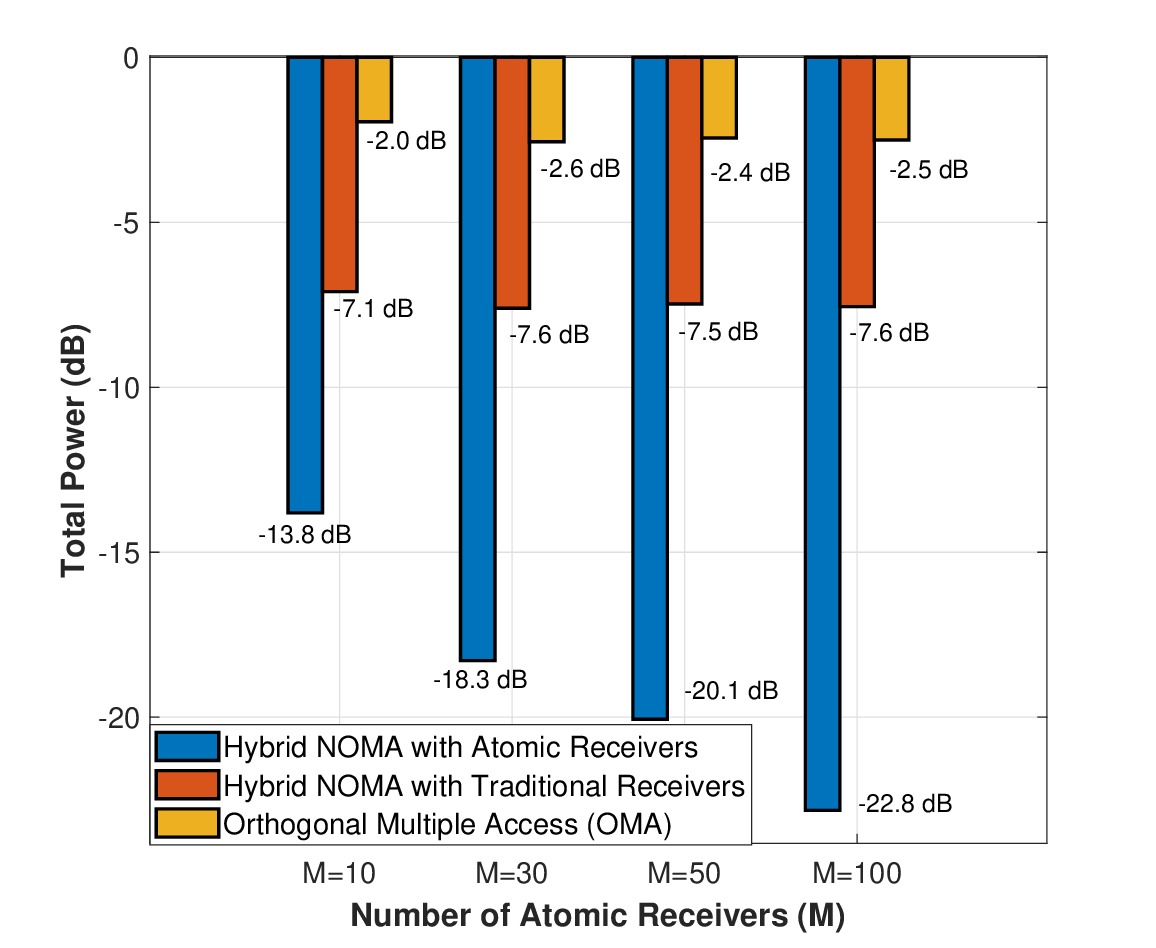}
	\par\vspace{2mm}
	\centering\textbf{(b)} SNR = 15 dB
	\label{fig:15dB}
	\vspace{0.5cm} 
	\includegraphics[width=0.85\linewidth]{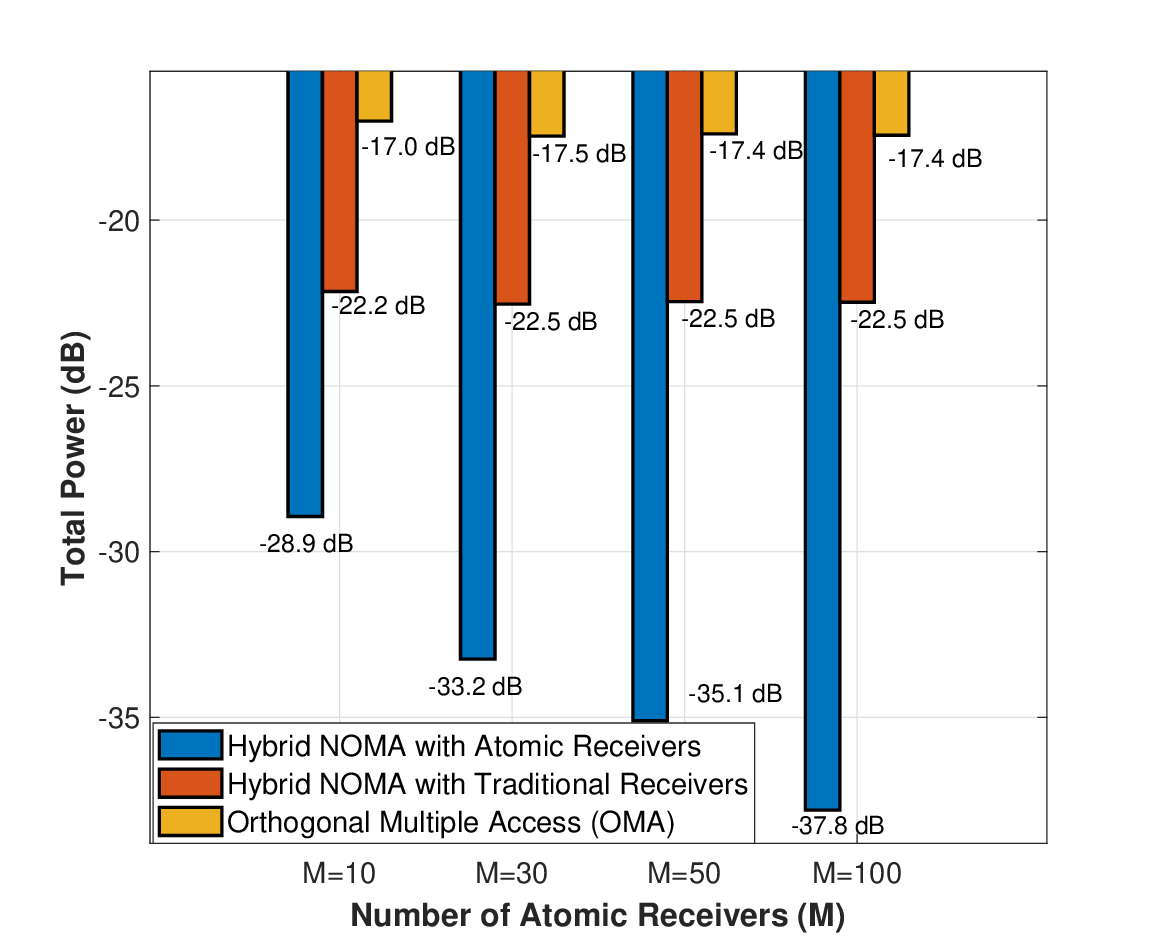}
	\par\vspace{2mm}
	\centering\textbf{(c)} SNR = 30 dB
	\label{fig:30dB}
	\caption{Power comparison of three transmission schemes for different $M$ values ($R=1.5$ bits/s/Hz).}
	\label{fig:Mzhuzhuang}
\end{figure}

A comprehensive bar chart comparison of the power consumption for the proposed scheme against the benchmarks (Traditional Receiver and OMA) is presented in Fig. \ref{fig:Mzhuzhuang}, spanning different SNR levels ($0$, $15$, and $30$ dB). A critical observation from these charts is that the power consumption of both Traditional-Receiver Hybrid NOMA and OMA baselines remains invariant with respect to $M$. In these benchmarks, we intentionally fix a single conventional RF front-end and do not scale the number of RF antennas with $M$; hence, they cannot exploit array or diversity gain without also incurring an $M$-fold increase in hardware and circuit power. By contrast, $M$ in the proposed RAQR architecture parameterizes the number of low-cost atomic sensing elements behind one RF port, so increasing $M$ enhances the effective channel gain without replicating RF chains, and the blue bars therefore exhibit a pronounced reduction in transmit power as $M$ grows. Consequently, the performance advantage of the proposed scheme becomes increasingly pronounced with larger $M$. Even at a relatively low SNR of 0 dB, scaling $M$ to 30 allows the proposed scheme to consume significantly less power than the benchmarks. This confirms that scaling the number of vapor cells is a highly effective strategy for overcoming adverse channel conditions and achieving superior energy efficiency compared to conventional approaches.

\begin{figure}[!htbp]
	\centering
	\includegraphics[width=0.85\linewidth]{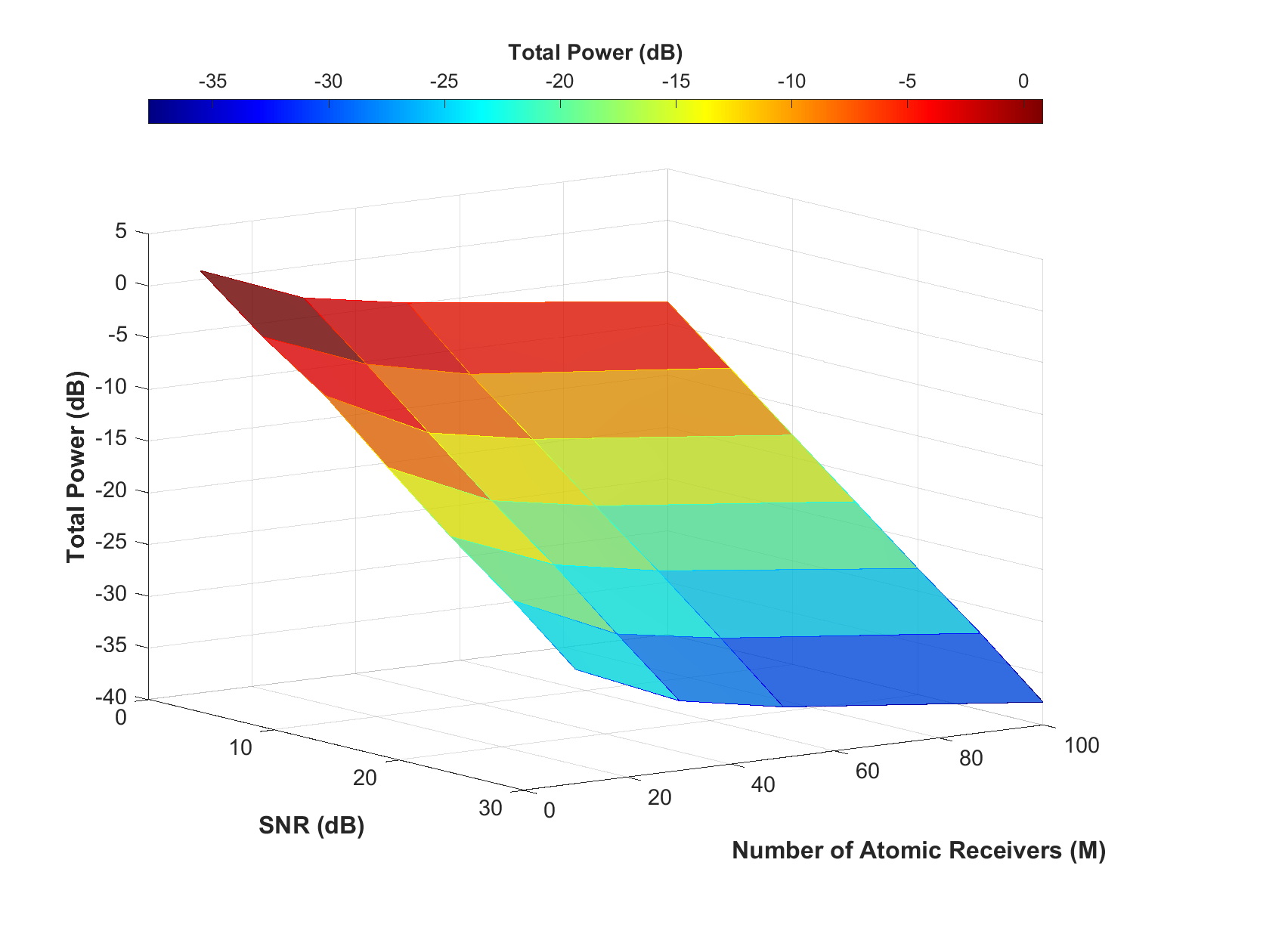}
	\caption{3D Surface: Total Power vs. SNR vs. $M$ for Hybrid NOMA with Atomic Receivers.}
	\label{fig:M3D}
\end{figure}
To visualize the joint impact of SNR and $M$ on the total system power, Fig. \ref{fig:M3D} provides a 3D surface plot. The curvature of the 3D surface visually reinforces the conclusions drawn from the 2D plots. The surface exhibits a steep gradient in the region where $M$ is small (10 to 30), particularly in low SNR regimes. This steep descent confirms that increasing the number of vapor cells is the most effective lever for reducing power consumption when signal quality is poor. As $M$ increases further, the surface flattens out, forming a "plateau" of high energy efficiency. This graphical representation succinctly summarizes the interplay between channel conditions and hardware resources, confirming that the proposed RAQR architecture can flexibly trade off array size for transmit power savings.

\subsection{Detailed Power Allocation Analysis}
\begin{figure}[!htbp]
	\centering
	
	\includegraphics[width=0.85\linewidth]{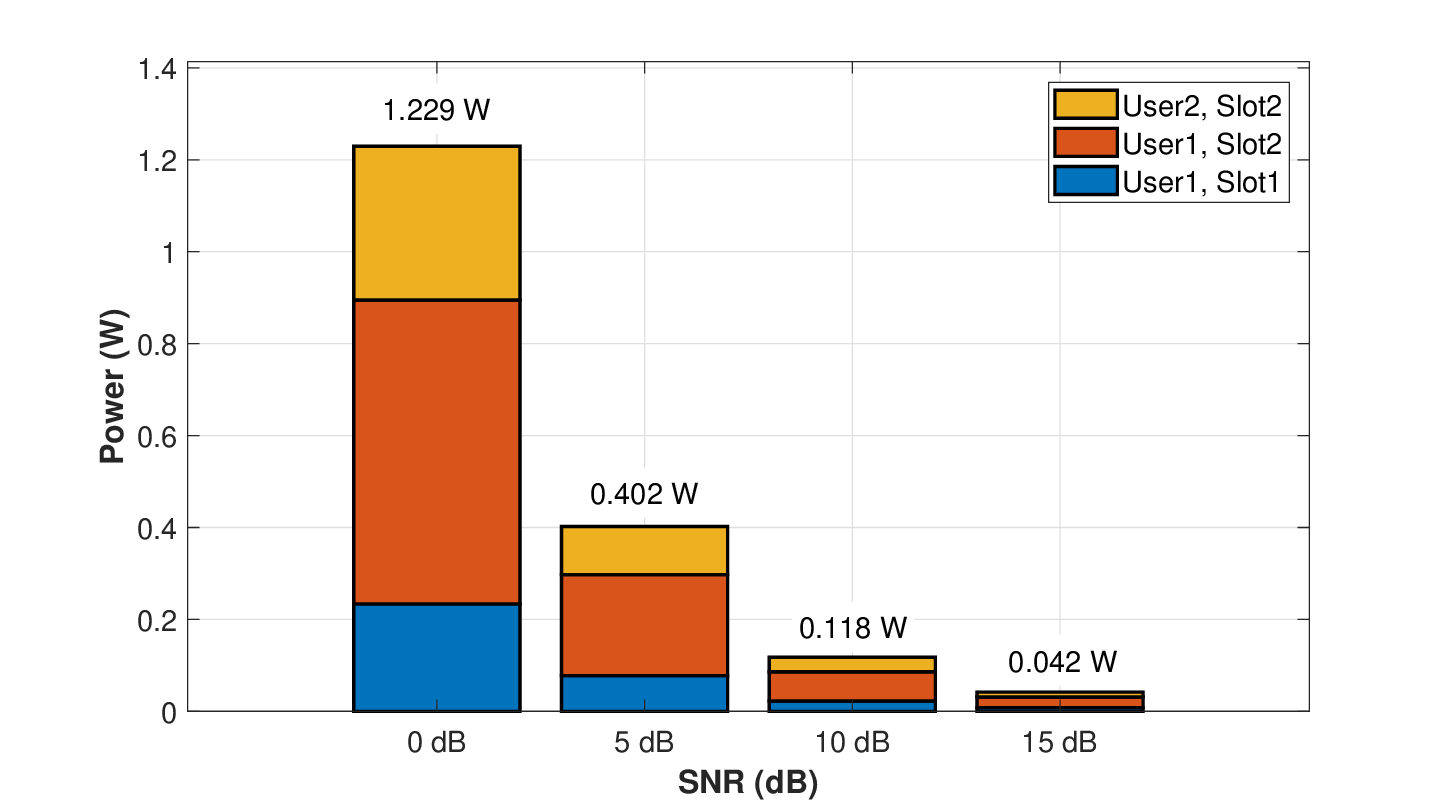}
	\par\vspace{2mm}
	\centering\textbf{(a)} Hybrid NOMA with Atomic Receivers
	\label{fig: Stacked Power Allocation_atomic receivers}
	
	\vspace{0.5cm} 
	
	\includegraphics[width=0.85\linewidth]{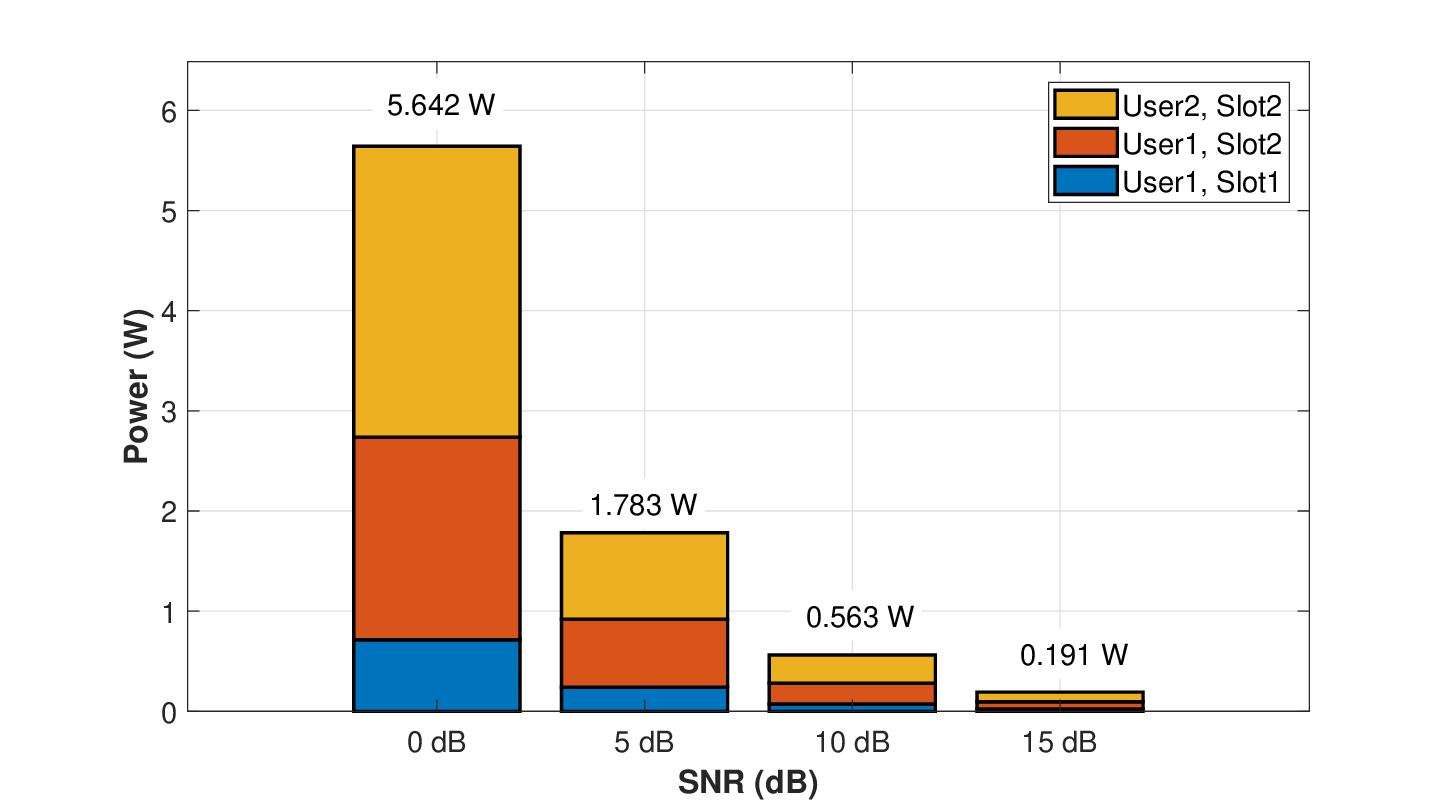}
	\par\vspace{2mm}
	\centering\textbf{(b)} Hybrid NOMA with Traditional Receivers
	\label{fig:Stacked Power Allocation_traditional receivers}
	
	\vspace{0.5cm} 
	
	\includegraphics[width=0.85\linewidth]{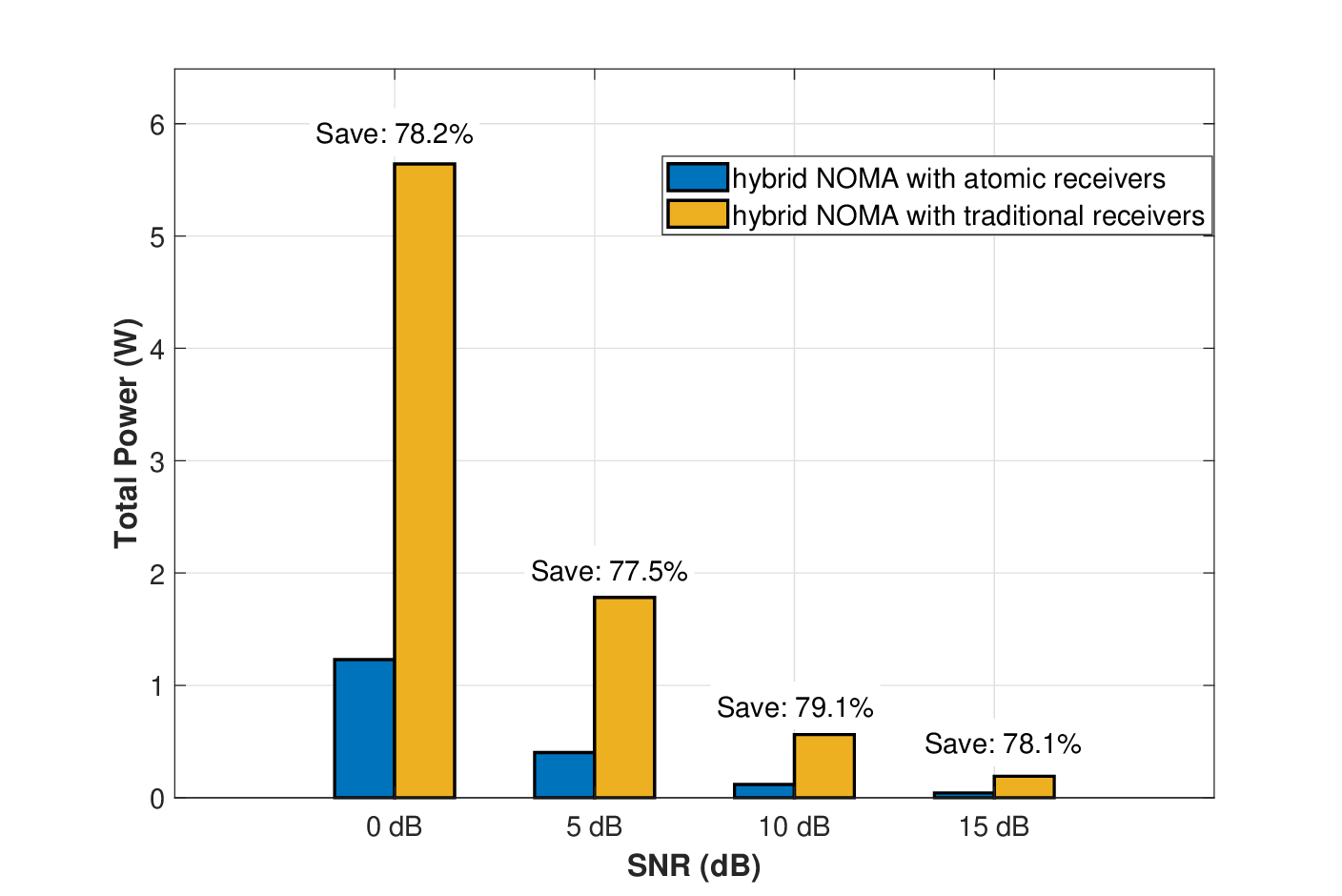}
	\par\vspace{2mm}
	\centering\textbf{(c)} Total Power Comparison 
	\label{fig:Stacked Power Allocation_Total Power Comparison}
	
	\caption{Detailed Power Allocation Analysis: Component-wise Breakdown.}
	\label{fig:power_allocation_analysis}
\end{figure}

To gain deeper insights into the internal mechanism driving the efficiency gains, we present a component-wise breakdown of power allocation in Fig.~\ref{fig:power_allocation_analysis}. This figure dissects the power distributed to User~1 (across Slots~1 and~2) and User~2 (in Slot~2) for both the proposed and benchmark schemes.

As shown in subplots~(a) and~(b), the Traditional-Receiver scheme typically demands substantially higher transmit power, particularly for the interference-limited components, to satisfy the decoding constraints (SINR thresholds). In stark contrast, the RAQR scheme significantly reduces the power required across all signal components. Subplot~(c) quantifies these savings, demonstrating a consistent and substantial reduction in total power consumption.

This reduction primarily reflects the coherent array gain provided by the $M$-element atomic aperture: by performing precise phase compensation and combining the responses of many vapor cells, the receiver enhances the effective channel gain, thereby allowing the transmitter to meet the same SINR targets with much lower energy. In this sense, the observed power gain is analogous to classical multi-antenna array gain; exploiting deeper quantum limits is left for future work.
\begin{figure}[!htbp]
	\centering
	\includegraphics[width=0.85\linewidth]{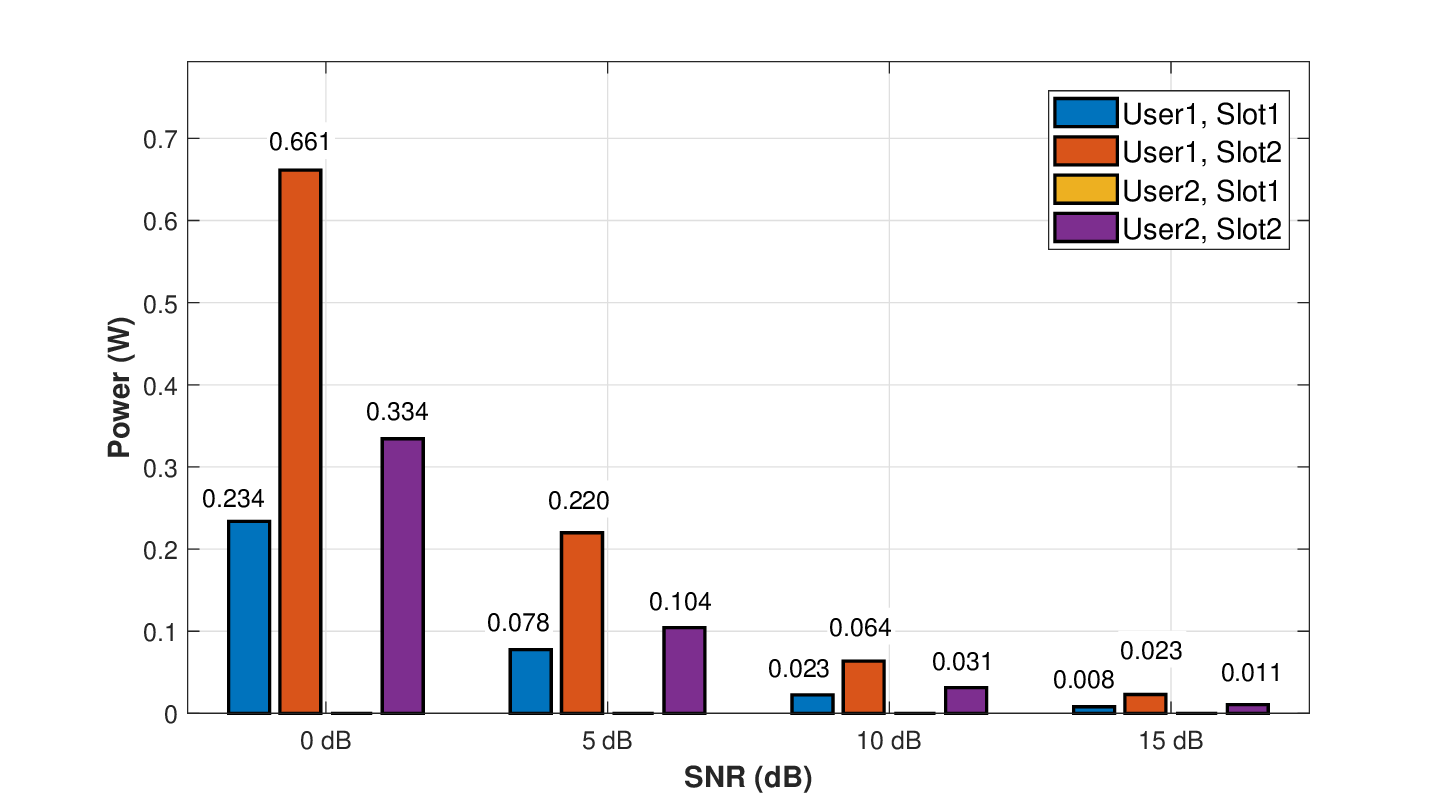}
	\par\vspace{2mm}
	\centering\textbf{(a)} Hybrid NOMA with Atomic Receivers - Power Allocation 
	\label{fig:hybrid NOMA with atomic receivers - Power Allocation}
	\vspace{0.5cm} 
	\includegraphics[width=0.85\linewidth]{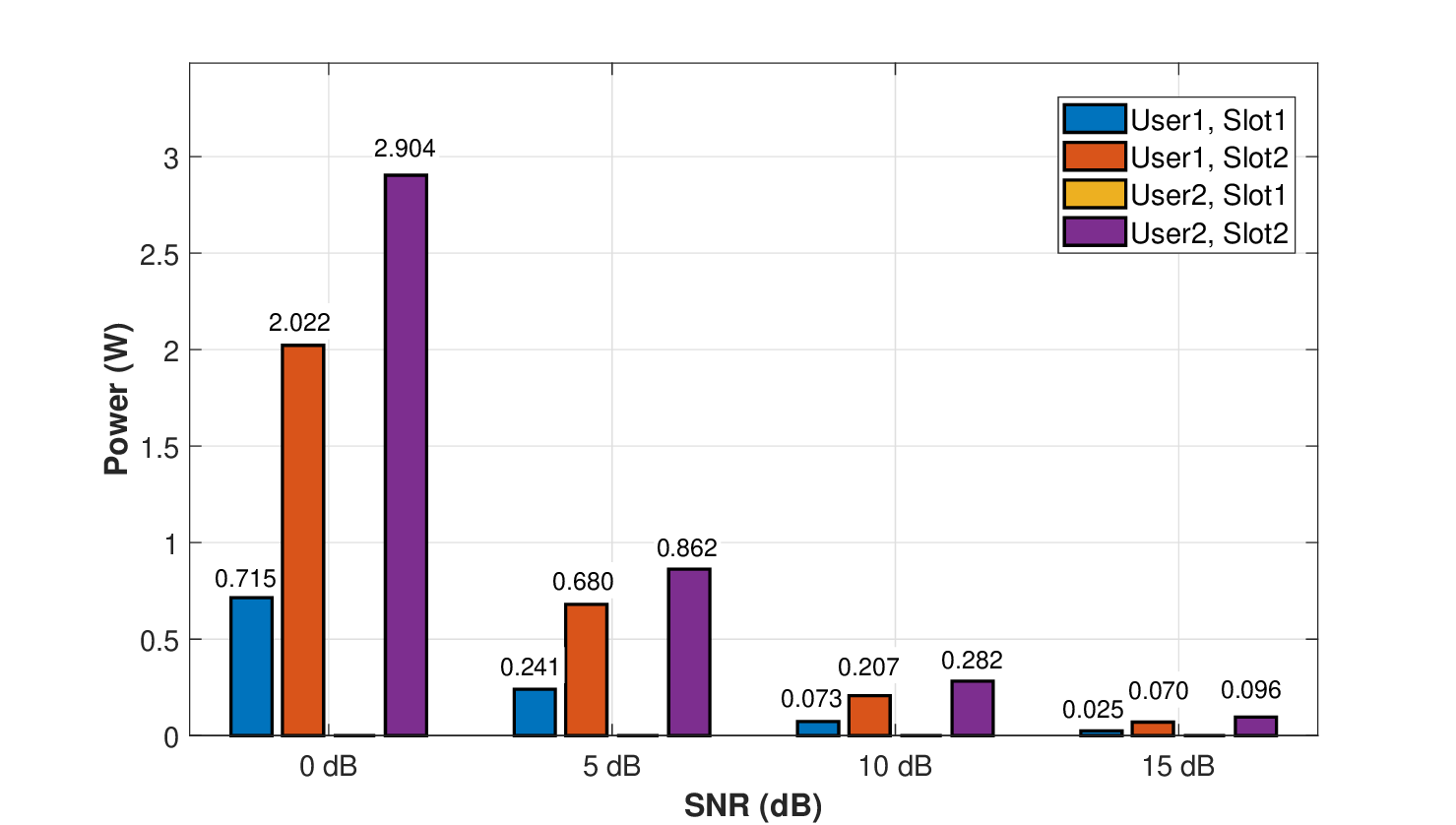}
	\par\vspace{2mm}
	\centering\textbf{(b)} Hybrid NOMA with Traditional Receivers - Power Allocation 
	\label{fig:hybrid NOMA with traditional receivers - Power Allocation}
	\caption{Power Allocation Comparison for Different Receivers ($R=1.5$ bits/s/Hz).}
	\label{fig:Linear vs logarithmic views of power allocation (R = 1.5)}
\end{figure}

Furthermore, as depicted in Fig. \ref{fig:Linear vs logarithmic views of power allocation (R = 1.5)}, a linear view of the allocation profile is offered to highlight structural differences. The scale reveals that the RAQR achieves a more concentrated and balanced power distribution, characterized by smaller dynamic range gaps between different user components. Conversely, the traditional receiver exhibits a large dynamic range, implying that excessive power is wasted compensating for receiver-side limitations in specific time slots. This indicates that the RAQR improves not only the total power consumption but also the structural efficiency of resource allocation, leading to a more stable and balanced network operation.

\subsection{Benchmarking against OMA and Robustness}
Finally, we benchmark the proposed scheme against standard baselines and assess its robustness under varying channel statistics.
\begin{figure}[!htbp]
	\centering
	\includegraphics[width=0.85\linewidth]{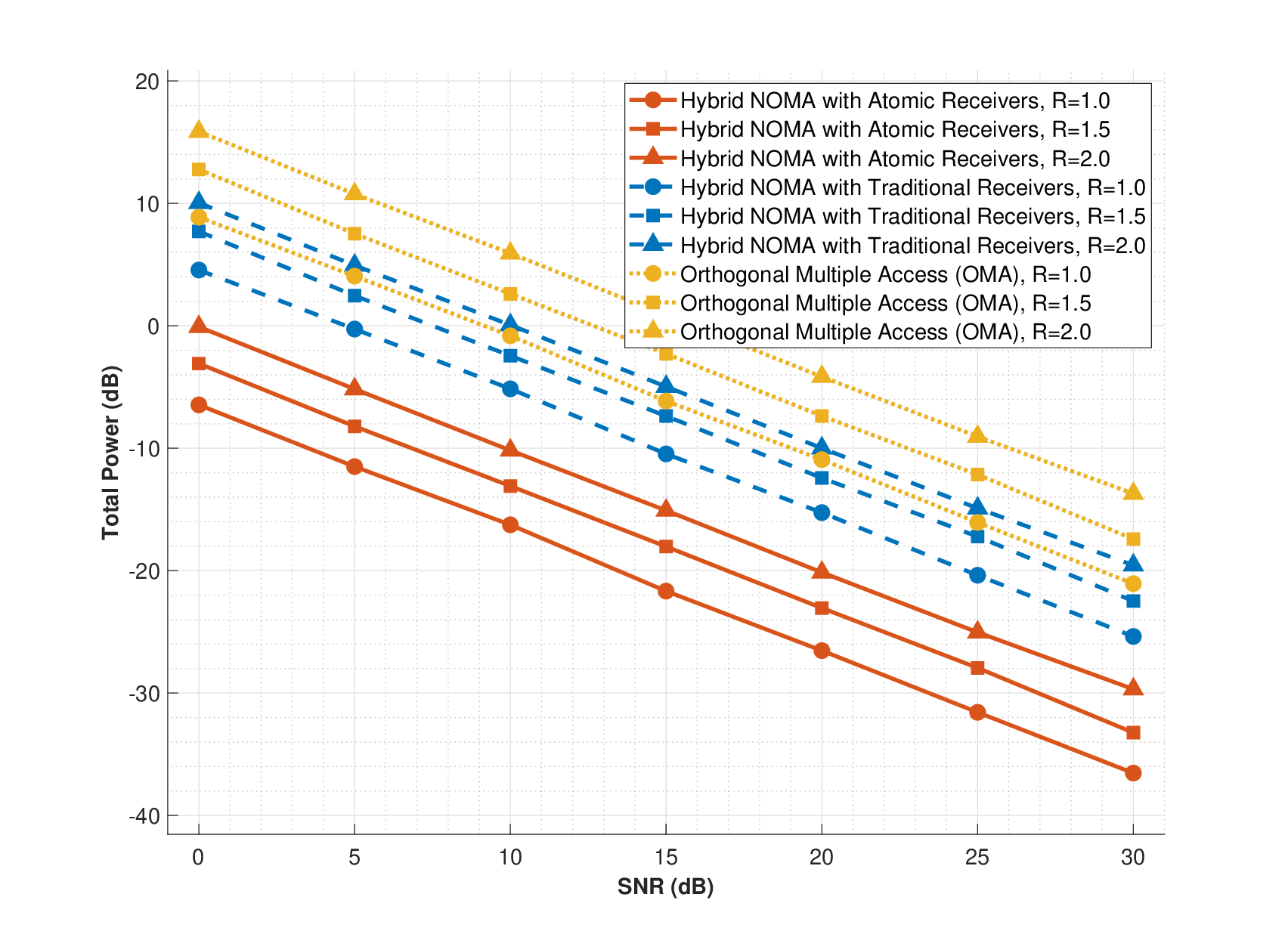}
	\caption{Total Power vs. SNR for Three Transmission Schemes ($M=10$).}
	\label{fig:m10}
\end{figure}
\begin{figure}[!htbp]
	\centering
	\includegraphics[width=0.85\linewidth]{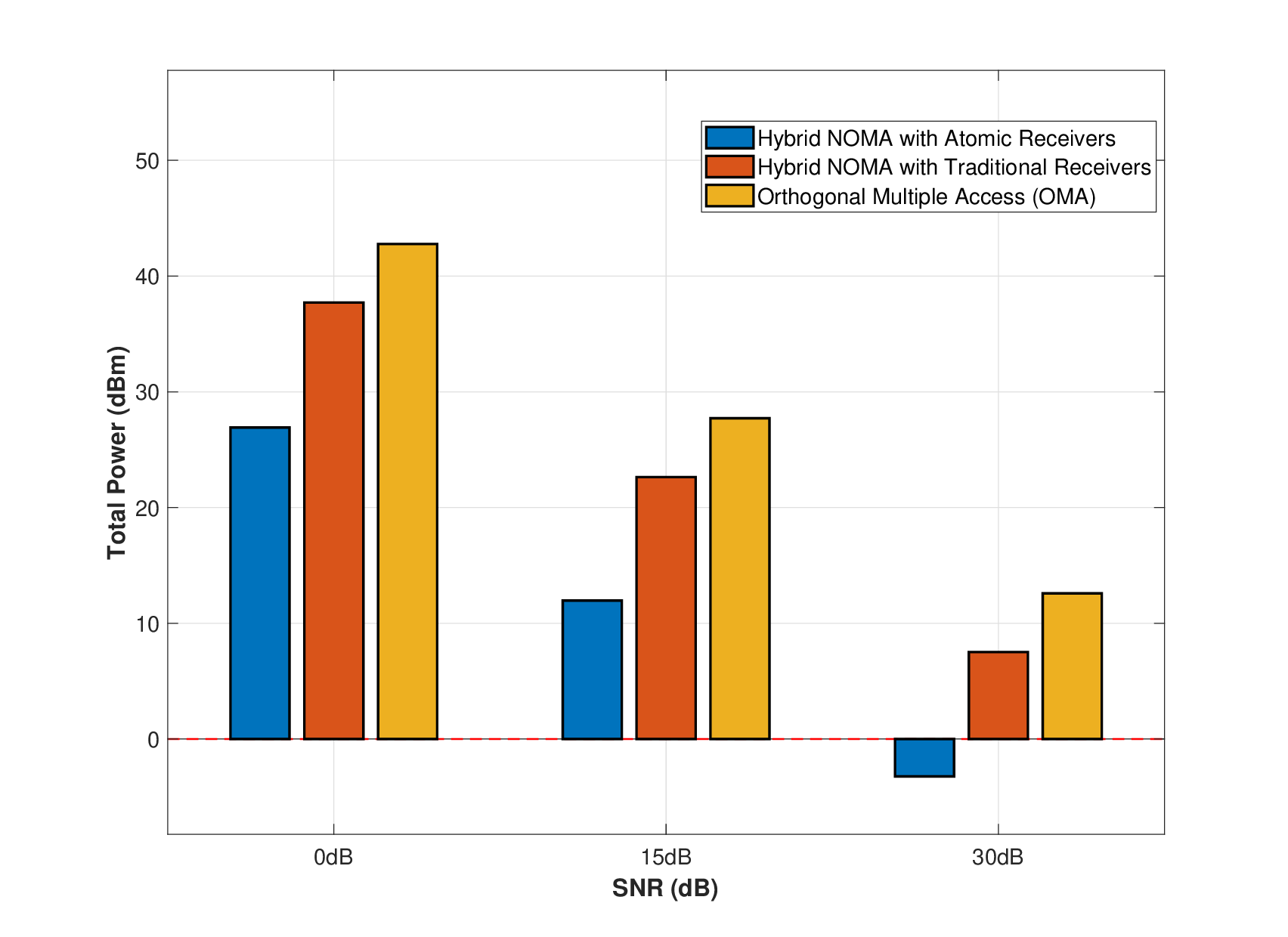}
	\caption{Total Power Comparison for Different Schemes ($R=1.5$ bits/s/Hz).}
	\label{fig:total power differences among the three transport strategies}
\end{figure}

A comparison of the three transmission strategies across the SNR spectrum is shown in Fig. \ref{fig:m10}, while Fig. \ref{fig:total power differences among the three transport strategies} provides a specific numerical snapshot at a target rate of $R=1.5$ bits/s/Hz.

The results clearly establish the superiority of the RAQR-Hybrid NOMA scheme across all target rates. As shown in the bar chart, the RAQR achieves an approximate 93\% power reduction compared to the OMA baseline. In comparison, the traditional receiver achieves only a 69\% reduction. This significant margin underscores the transformative potential of integrating atomic sensing into NOMA networks.
\begin{figure}[!htbp]
	\centering
	\includegraphics[width=0.85\linewidth]{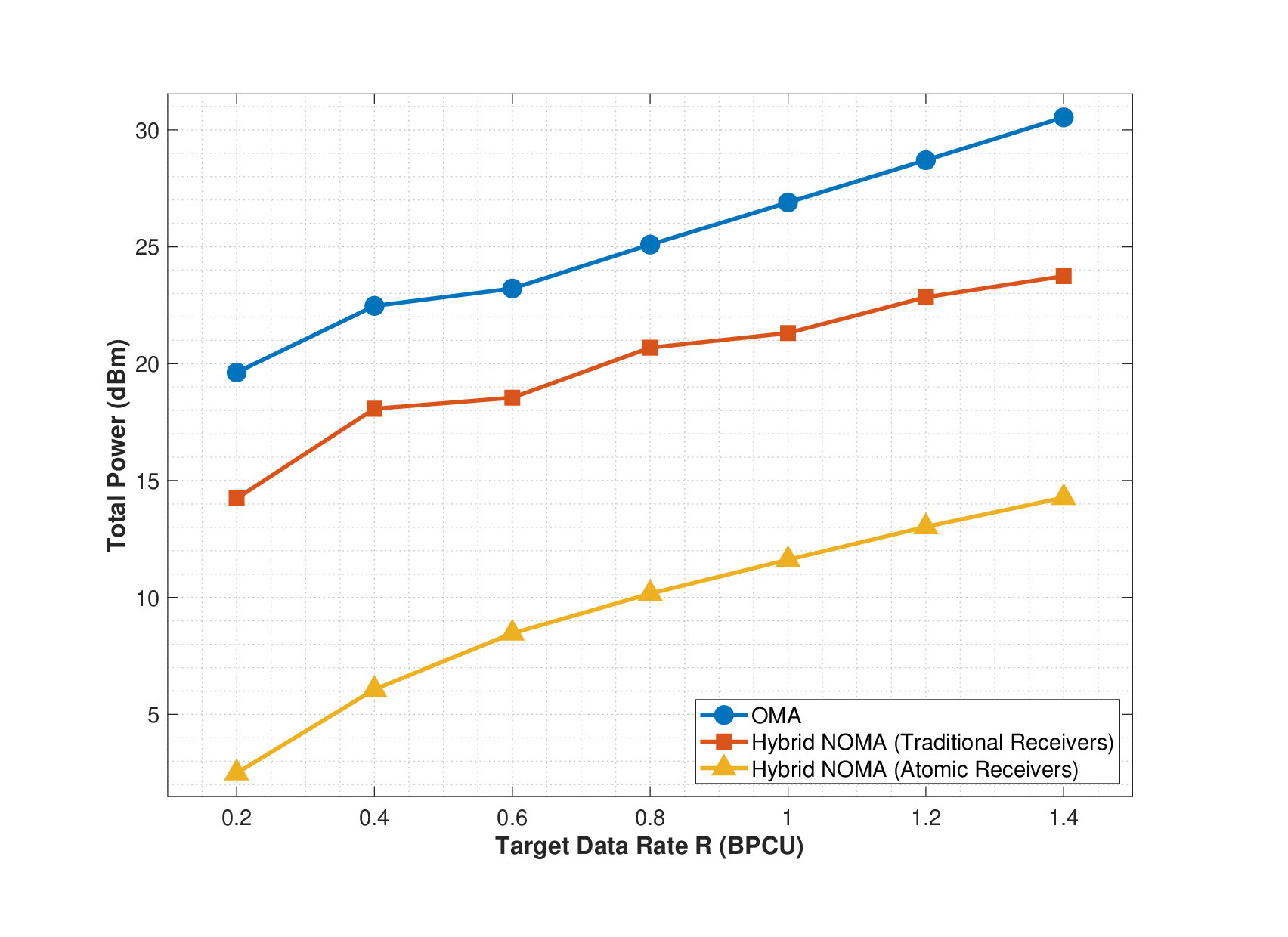}
	\par\vspace{2mm}
	\centering\textbf{(a)} Case I: Symmetric Average Signal Strength 
	\label{fig:Case I}
	\vspace{0.5cm} 
	\includegraphics[width=0.85\linewidth]{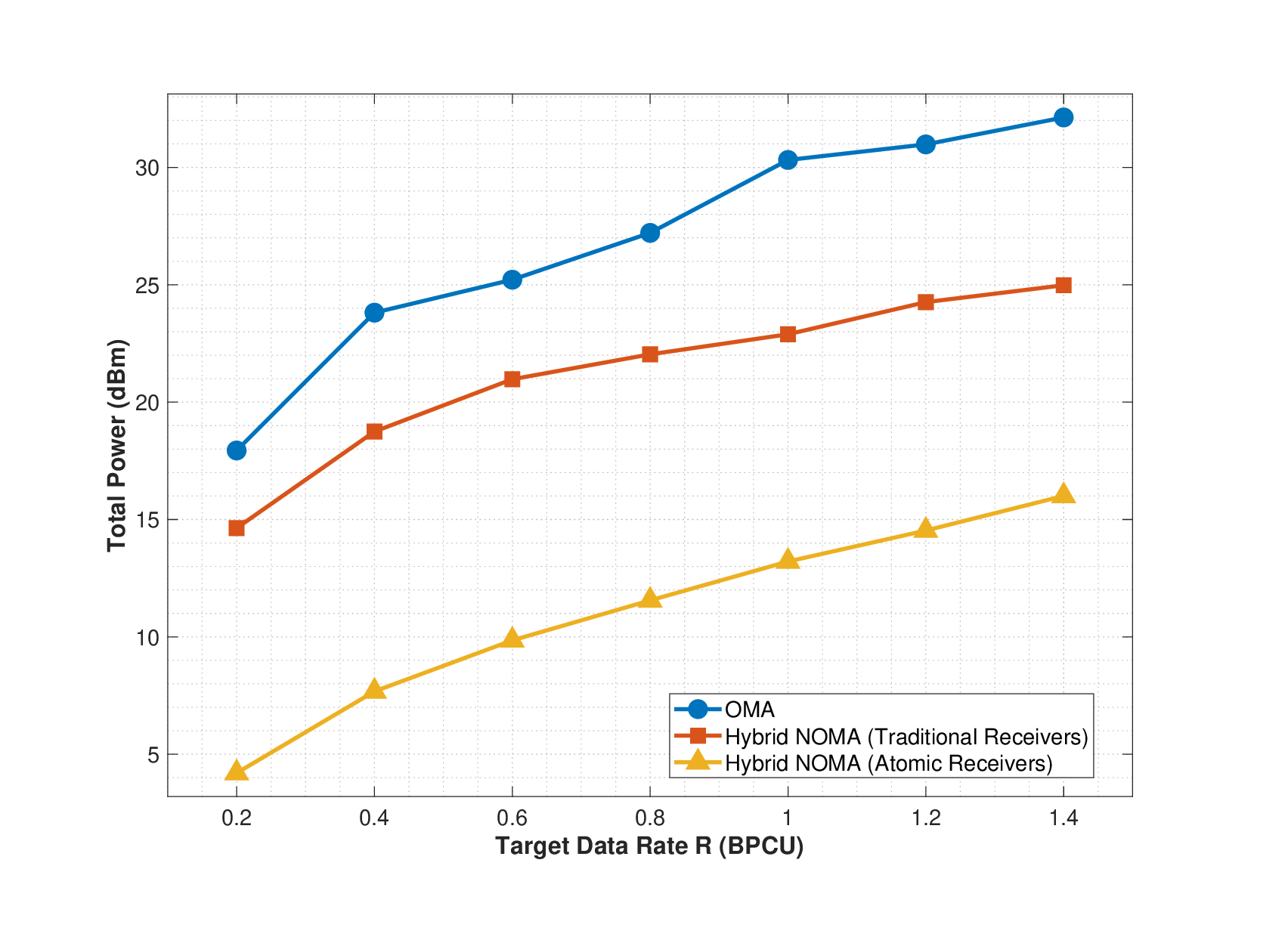}
	\par\vspace{2mm}
	\centering\textbf{(b)} Case II: Asymmetric Average Signal Strength 
	\label{fig:Case II}
	\caption{Total Power Consumption vs. Target Data Rate under Different Channel Cases.}
	\label{fig:target rate}
\end{figure}

Figure.~\ref{fig:target rate} evaluates the total transmit power versus target rate $R$ under symmetric (Case~I) and asymmetric (Case~II) channel conditions. The proposed RAQR-Hybrid NOMA consistently achieves the lowest power consumption, followed by traditional Hybrid NOMA while OMA requires the highest power. This underscores the robustness of the atomic architecture regardless of channel statistics.\\
\indent Beyond energy efficiency, the figure also reveals the spectral efficiency advantage of NOMA: for a given power budget, RAQR-Hybrid NOMA supports a higher $R$ than its counterparts, as evidenced by its slower power growth with $R$. Moreover, when combined with the power-allocation profiles in Fig.~\ref{fig:Linear vs logarithmic views of power allocation (R = 1.5)}, the RAQR design enhances user fairness by yielding a more balanced power distribution across users. This is because the atomic receiver compensates for channel disparities through array gain and phase alignment, reducing the need for excessive power to weak users. These benefits hold under both symmetric and asymmetric channels, confirming that the proposed framework simultaneously improves energy efficiency, spectral utilization, and fairness.

\section{Conclusion and Future Work}\label{section:Conclusion}
In this paper, we have proposed an innovative three-layer decomposition optimization framework for uplink Hybrid NOMA systems empowered by RAQR. By rigorously modeling the physical constraints of atomic sensing, specifically the non-linear amplitude-only observation, we successfully decoupled the intractable joint optimization problem into three mathematically tractable sub-modules. Specifically, the inner layer addresses fundamental phase ambiguity using a customized EM-GS algorithm to jointly estimate user symbols and phase compensation parameters. Subsequently, the middle layer derives a closed-form power allocation solution based on KKT conditions to guarantee optimal energy efficiency and finally, the outer layer employs linear MMSE detection for high-reliability signal recovery.

The results clearly demonstrates the superiority of the RAQR-Hybrid NOMA scheme across all target rates. The RAQR achieves an approximate 93\% power reduction compared to the OMA baseline, whereas the traditional receiver achieves only a 69\% reduction. This significant margin confirms the energy-efficiency synergy of the proposed architecture. However, investigating whether this high sensitivity can further mitigate NOMA-specific impairments, such as SIC error propagation or stringent power imbalance constraints, remains a challenging open issue for future research.


\balance
\bibliographystyle{ieeetr}
\bibliography{my_references}

\end{document}